\documentclass[3p]{elsarticle}

\usepackage{graphicx}
\usepackage{psfrag}
\date{today}
\usepackage{xcolor}
\graphicspath{{fig_physD/}}

\begin{document}

\begin{frontmatter}


\title{
Decaying turbulence: what happens when the correlation length varies spatially in two adjacent zones.
}
\author{Daniela Tordella* and Michele Iovieno\\
{\itshape
Dipartimento di Ingegneria Aeronautica e spaziale, Politecnico di Torino,\\ Corso Duca degli Abruzzi 24, 10129 Torino, Italy.\\
}}

\begin{abstract}

We have imagined a numerical experiment to explore the onset of turbulent intermittency  associated with a spatial perturbation of the correlation length.
We place two isotropic regions, with different integral scales, inside a volume where the turbulent kinetic energy is initially uniform and leave them to interact and evolve in time. The different length scales produce different decay rates in the two regions. Since the smaller-scale region decays faster, a transient turbulent  energy gradient is generated at the interface between the two regions. The transient is characterized by three  phases in which the kinetic energy gradient across the interface grows, peaks and then slowly decays. The transient lifetime is almost proportional to the initial ratio of the correlation lengths. The direct numerical simulations also show that the interface width grows in time. The velocity moments inside this interaction zone are seen to depart from their initial isotropic values and, with a certain lag, the anisotropy is seen to spread to small scales. The longitudinal derivative moments also become anisotropic after a few eddy turnover times.  This anisotropic behaviour is different from that observed in sheared homogeneous turbulent flows, where high transverse derivative moments are generated, but longitudinal moments almost maintain  the  isotropic turbulence values.
Apart from the behaviour of the energy gradient transients, the results also show the timescaling of the interface diffusion width, and data on the anisotropy of the large and small scales, observed through one-point statistics determined inside the intermittency sublayer, which is associated with the interaction zone.
\end{abstract}

\end{frontmatter}

\section{Introduction}

In recent years, the Kolmogorov postulate on local isotropy (PLI) has been examined by means of laboratory and numerical experiments (\cite{antonia84}--\cite{gw98}).  The current phenomenological interpretation of real turbulent flows is based on this postulate, which maintains that turbulence in fluids can achieve a universal state at small scales where fluctuations are supposed to become statistically isotropic, and as a consequence universal. The contravention of this postulate implies that small scales in a turbulent system interact directly with larger scales. Thus, a sort of long-term interaction occurs and is able to induce anisotropy on the small scales, according to the symmetry properties of the system. Direct communication
between large and small scales at the same time undermines the
concept of small-scale universality and that of energy cascade.
The problem remains on how to examine the effects of
large-scale anisotropy on small scales. The answer is to examine statistics that
are sensitive to flow anisotropy.

The main way of verifying the PLI is based on turbulence studies at very high Reynolds numbers
($Re \sim 10^4$ and greater). For example, a popular way of measuring small-scales anisotropies is to obtain the $Re$ dependence of the isotropic and anisotropic statistical observables on the basis of velocity gradients.

\begin{figure}
 \centering
\includegraphics[width=0.56\textwidth]{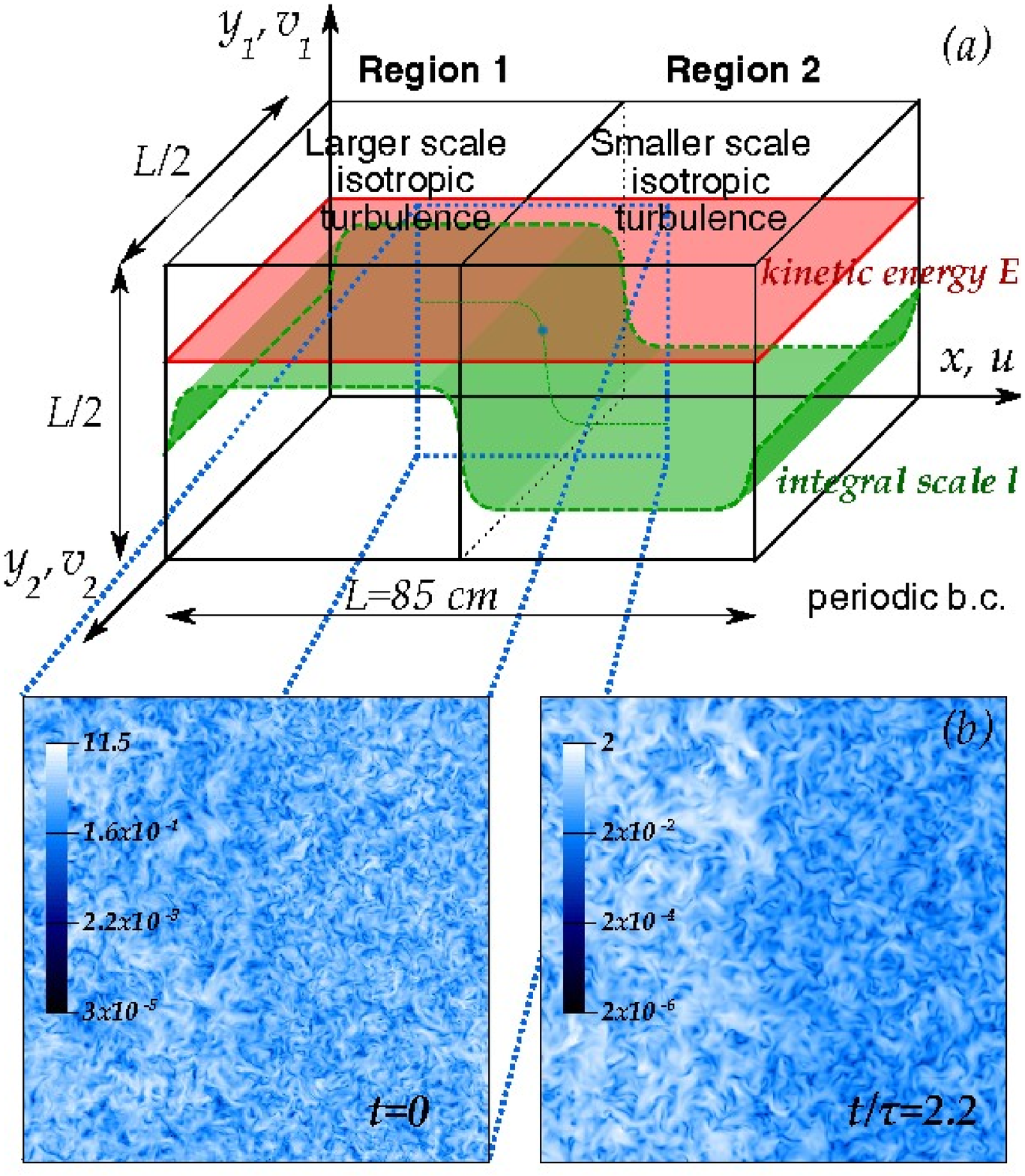}\\[4mm]
\includegraphics[width=0.56\textwidth]{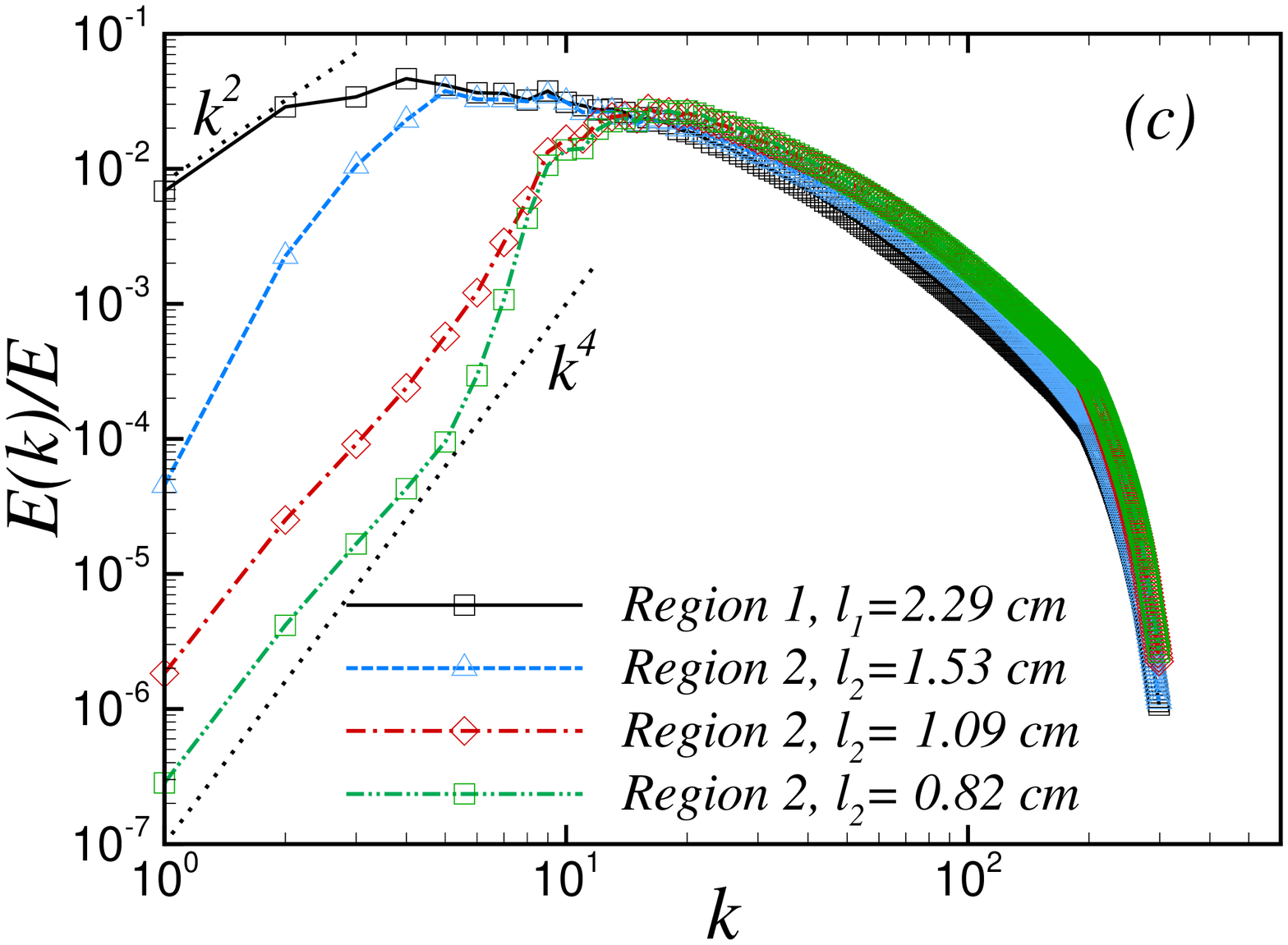}

\vskip -2mm
\caption{(a) Flow scheme. The Taylor microscale Reynolds number of the turbulence in Region 1 is 150. By varying the correlation length from 0.8 to 2.8 cm, it varies from 70 to 150 in region 2. Due to the use of cyclic boundary conditions, a second statistically equivalent interaction layer is produced at the border of the domain. (b) View of the energy profile in the central part of a plane $(x,y_1, y_2=const)$ at different instants.  The second interaction layer is not visible in these images. (c) Initial three-dimensional normalized spectra $E(k)/E$ in homogeneous regions 1 and 2. Dimensional reference data for the experiment:
velocity fluctuation $u'_{rms}= 0.86 m/s$, kinetic energy $E_1 = 1.1$ J/kg, molecular viscosity $\nu=1.4 \times 10^{-5}$ m$^2$/s, size of the flow domain 
$L=0.85$ m (dimension along x), $L/2=0.425$ m (dimension along $y_1$ and $y_2$),  correlation lengths: $\ell_1 = 2,29 \times 10^{-2}$ m, field 1, $\ell_2 = 1.53 \times 10^{-2}$ m, $1.09 \times 10^{-2}$ m; $8.2 \times 10^{-3}$ m; field 2, time scales: $\tau_1 = 0,027$ s, field 1, $\tau_2 = 0.018$ s, $0.012$ s, $0.0095$ s (for field 2 yielding, the scale ratio is: 2.8, 2.1,1.5, respectively).
  }
\label{fig.schema}
 \end{figure}

 \begin{figure}
  \centering
  \vskip -5mm
  \psfrag{b}{\large $a$}
  \includegraphics[width=0.49\textwidth]{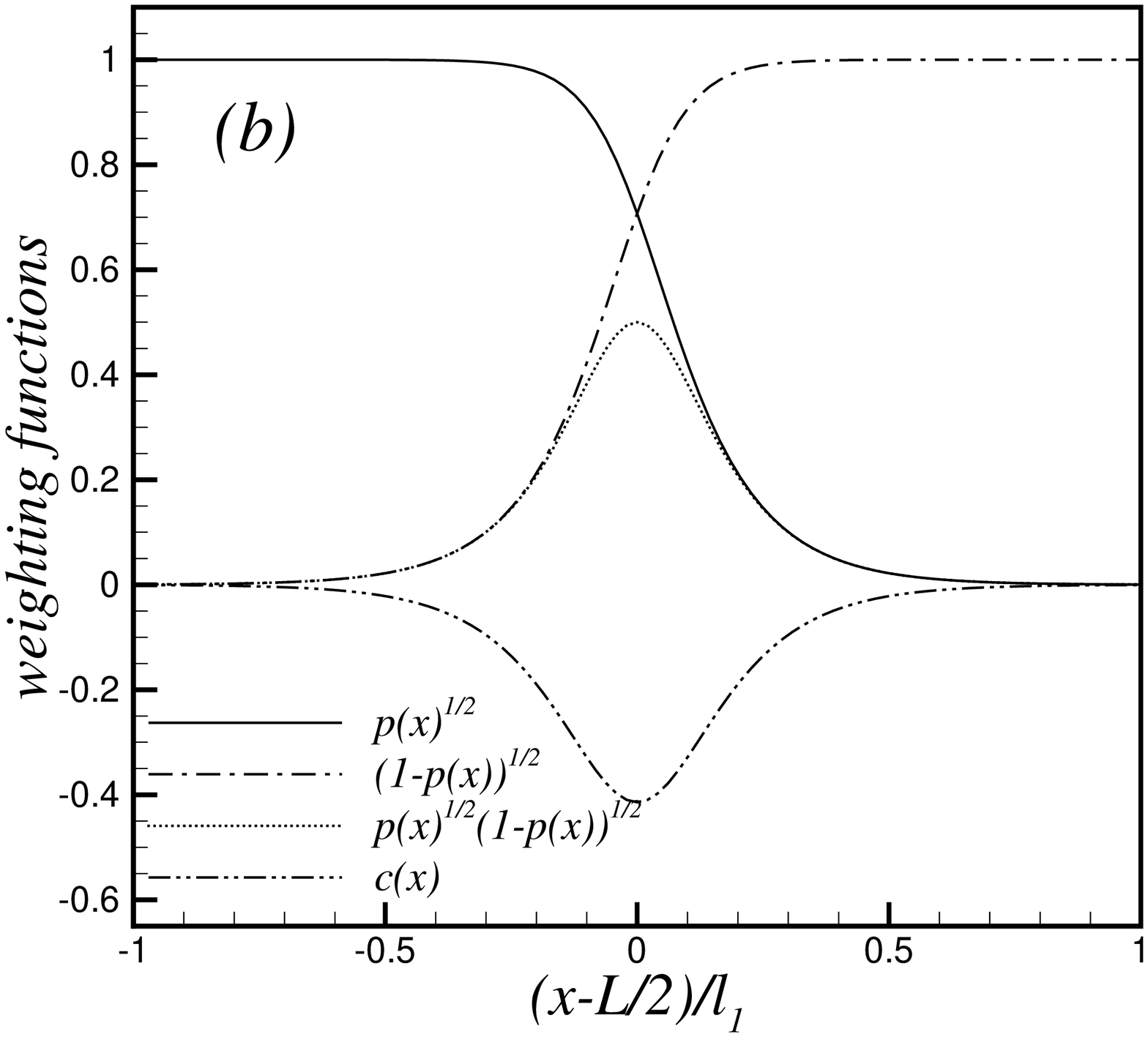}
  \psfrag{a}{\large $b$}
  \includegraphics[width=0.49\textwidth]{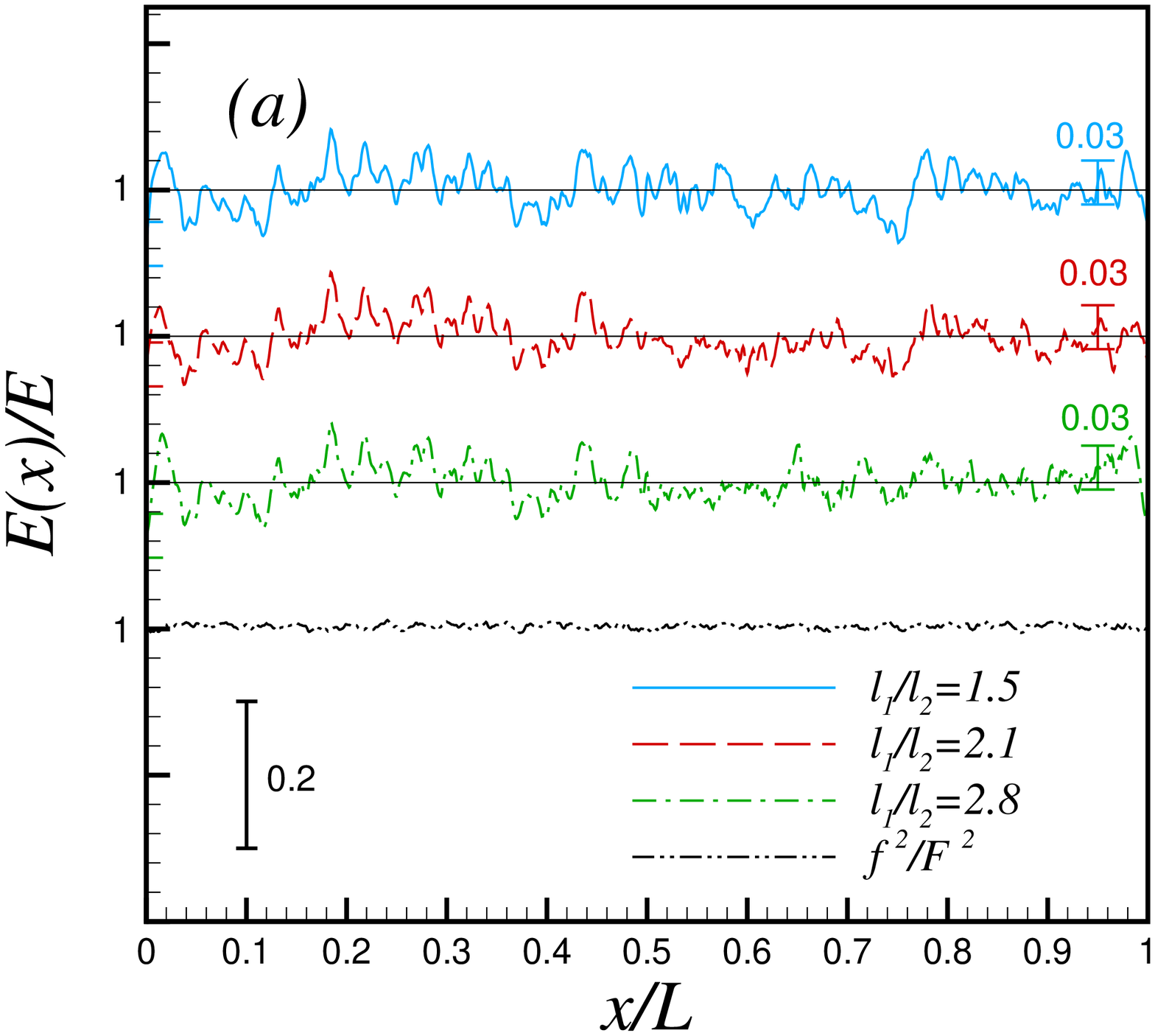}
  \vskip -1mm
  \caption{(a) Weighting functions used to generate the initial conditions in the central part of the domain.
The initial condition is generated as $u_i = u_i^{(1)}p(x)^{\frac{1}{2}} + u_i^{(2)}(1-p(x))^{\frac{1}{2}} + fc(x)$, where $u_i^{(1)}$ and $u_i^{(2)}$ are two homogeneous and isotropic fields with integral scales $\ell_1$ and $\ell_2$,
$p(x)=\left[ 1- \tanh a (x-L/2)/\ell_1) \right]/2$ is a smoothing function; $cf$ is a correction introduced to obtain a uniform energy distribution: $c=1-(p^{1/2}+(1-p)^{1/2})$ is a spatial modulation in the smoothing region and $f$ is a random field with zero mean and uniform variance $F^2$. The value of $F^2$ is chosen a posteriori in order to generate a uniform energy level $E$. The coefficient \textit{a} in function p(x) is equal to 11 if the matching variable is normalized on the correlation length $\ell_1$ ( $\tanh a (x-L/2/\ell_1) $) and is equal to 400 if the matching variable is normalized on $L$  ($\tanh a (x-L/2/L) $). (b) Initial distribution of the average turbulent kinetic energy. 
}
\label{fig.matching}
\end{figure}


The relationship between intermittency and small-scale anisotropy is
considered  to determine whether small-scale isotropy is
restored at very high Reynolds numbers. The set-up often used to test the return to isotropy is a homogeneous shear flow in which the large-scale
mean velocity has a linear profile. In many previous works (e.g. \cite{gw98, ws00,ws02,ssk03}), researchers looked at the increments in the streamwise ($x$ direction) fluctuating
velocity, $u$, in the direction of the mean shear $S_{\partial U/\partial y}$, where $y$ is the transverse direction and $U(y)$ is the mean
velocity. 

The effect of the Reynolds number on turbulence small-scale anisotropy and intermittency is considered
in an attempt to separate large and small scales, with respect to their  size and energy, in order to reduce the probability of mutual communication, 
Although Taylor scale Reynolds numbers of $10^3$ (equivalent to a turbulence Reynolds number based
on the integral scale of $10^5$) can be achieved in the laboratory, it is argued \cite{w09}
that there is still a need to conduct experiments and computations at even
higher Reynolds numbers to resolve the outstanding basic and applied issues in
turbulence.

In our opinion, this is not the only approach that can be used to test the PLI. Another approach is  to focus on
flows at a given Taylor scale Reynolds number and to slightly perturb the state of homogeneity and isotropy to see how and to what extent the perturbation is felt by the small scales during the transient. This method is inspired by the Initial Value Problem in stability theory \cite{drazin,stc09,stc10}, where the basic laminar flow is arbitrarily perturbed by an infinitesimal perturbation with a given wavenumber. The mildest perturbation that can be conceived without forcing the turbulence (i.e.\ without introducing a production of kinetic energy in the system) is that of varying the correlation length of the turbulence in the space, while avoiding the generation of a mean shear. As known, the coupling of shear with fluctuation is a source of kinetic energy.   

In order to achieve this situation, we consider a numerical experiment in which we follow the time decay of a turbulent flow with an initially uniform distribution of turbulent kinetic energy, but where the integral scale has been varied slightly in two adjacent regions.
The two regions are characterized by different spectrum shapes in the low wavenumber range, one almost $k^2$ and the other close to $k^4$ \cite{s67,kd06}. Thus, we have a condition where we can observe a different decay generated by the scale inhomogeneity. 
This simple configuration allows us to investigate the generation of large-scale and small-scale anisotropy arising from the different decay rates of the two interacting isotropic fields, \cite{g92}--\cite{a07}. The  different decay exponents shown by the two fields generate an energy step, and thus a layer where the flow is neither homogeneous nor isotropic. Thus, the system is characterized by the simultaneous presence of a scale and an energy gradient. No mean velocity gradient is present. No preferential direction is induced by the mean flow gradient \cite{euromech512,ti09}. 

The paper is organized as follows. Section 2 presents the DNS method, the simulation conditions and the turbulence characteristics. Section 3 is devoted to the general lineaments of the temporal evolution of the interaction layer. Section 4 presents an analysis of the transient lifetime, in which the onset, growth, peaking and decay of anisotropy and of the kinetic energy gradient can be observed. Scaling laws that link the velocity and velocity derivative skewness and kurtosis  are proposed, on the basis of best fitting of the results, produced by three simulations  parametrized by the ratio of the correlation lengths of the two isotropic interacting fields ($\ell_1/\ell_2 = 1.5, 2.1, 2.8$). We also discuss the difference with the anisotropy present in homogeneous shear flows. The Reynolds number, based on the Taylor microscale of the interacting fields, varies from 70 to 150, when the correlation length is varied from 0.8 to 2.8 cm. Section 5 provides the conclusions.



\addtolength{\textheight}{1cm}
\begin{figure}
 \centering
  \includegraphics[width=0.60\textwidth]{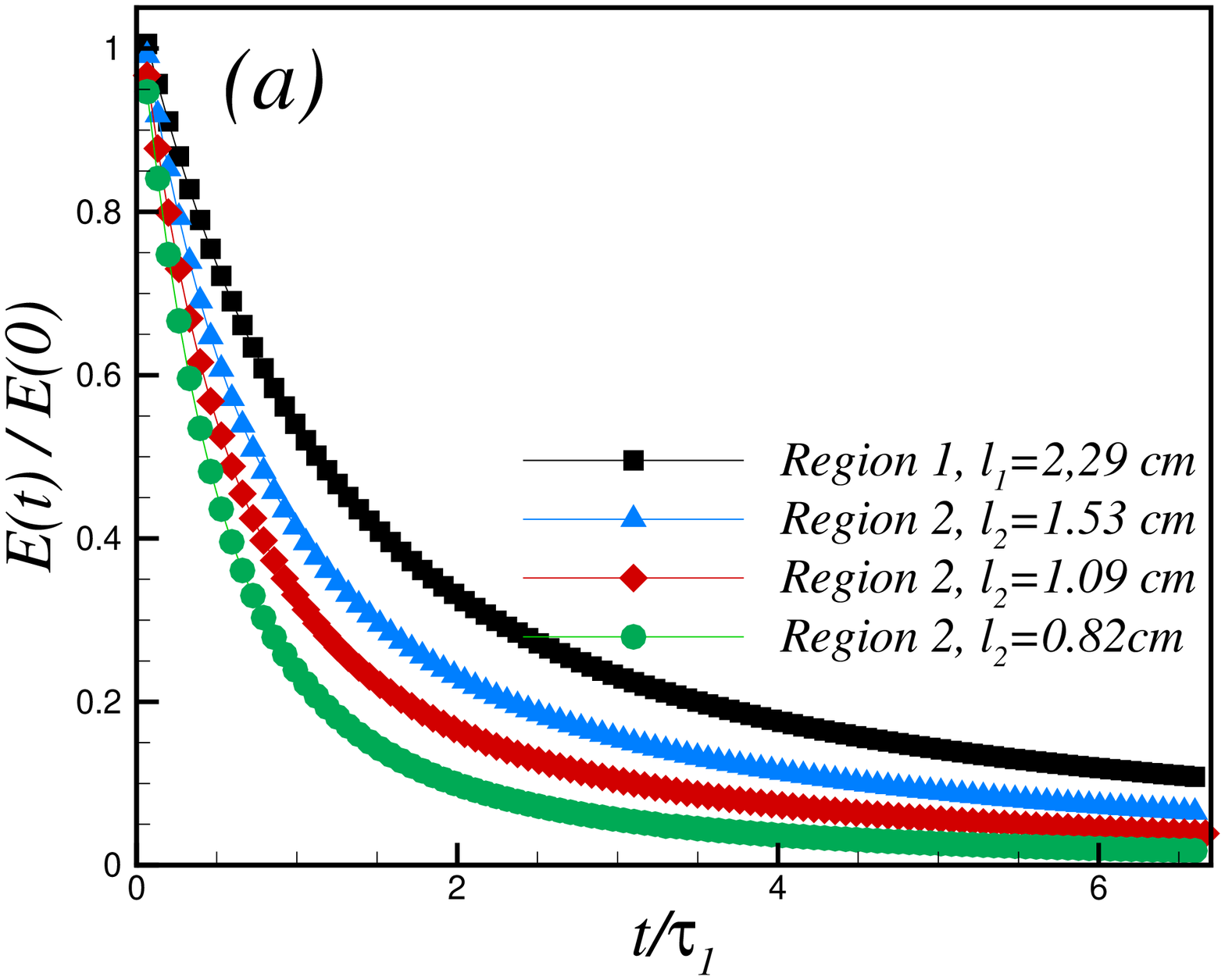}\\
  \vspace*{-7.8cm}\hspace*{0.42\textwidth}\includegraphics[width=0.30\textwidth]{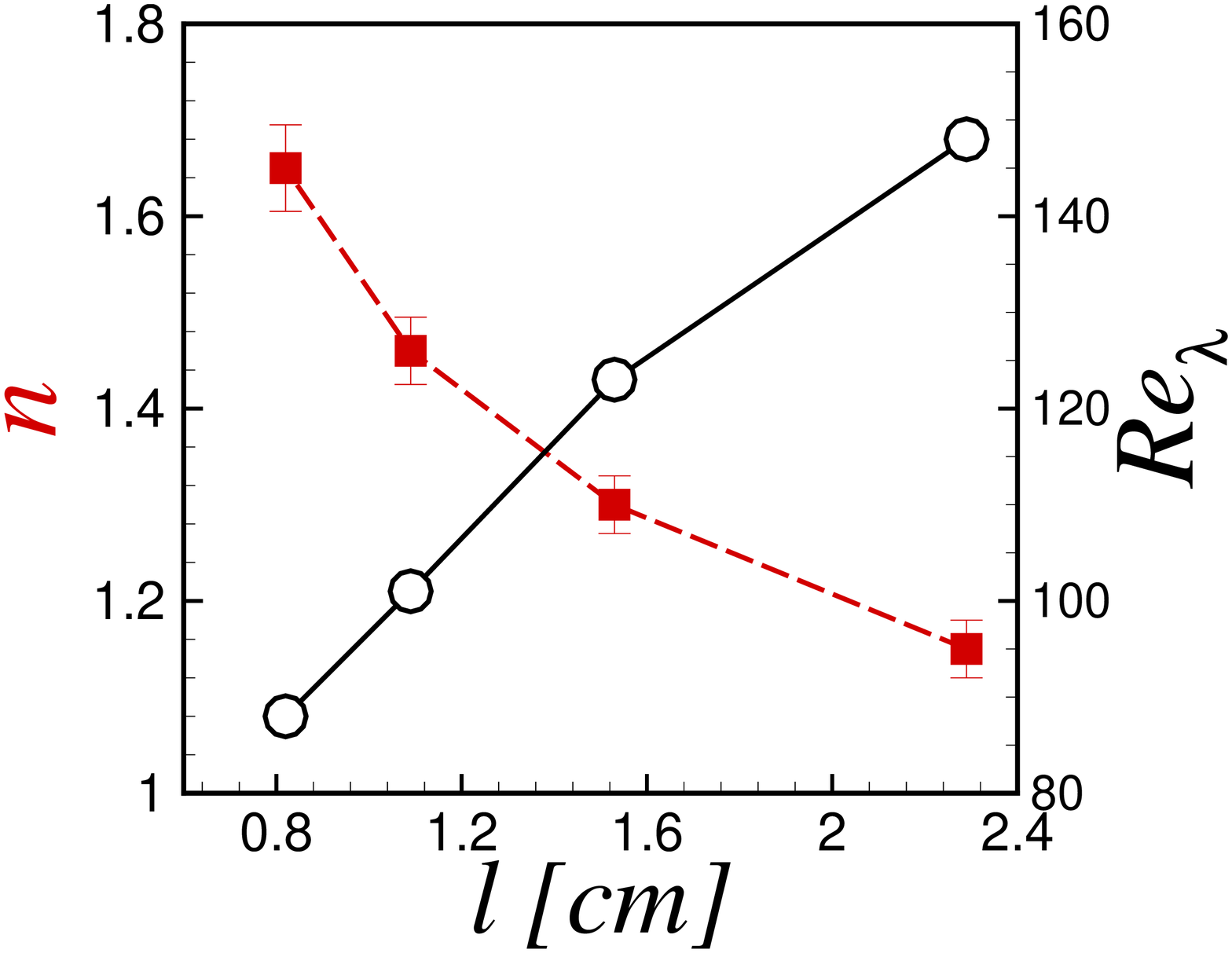}\\[5cm]
  \includegraphics[width=0.60\textwidth]{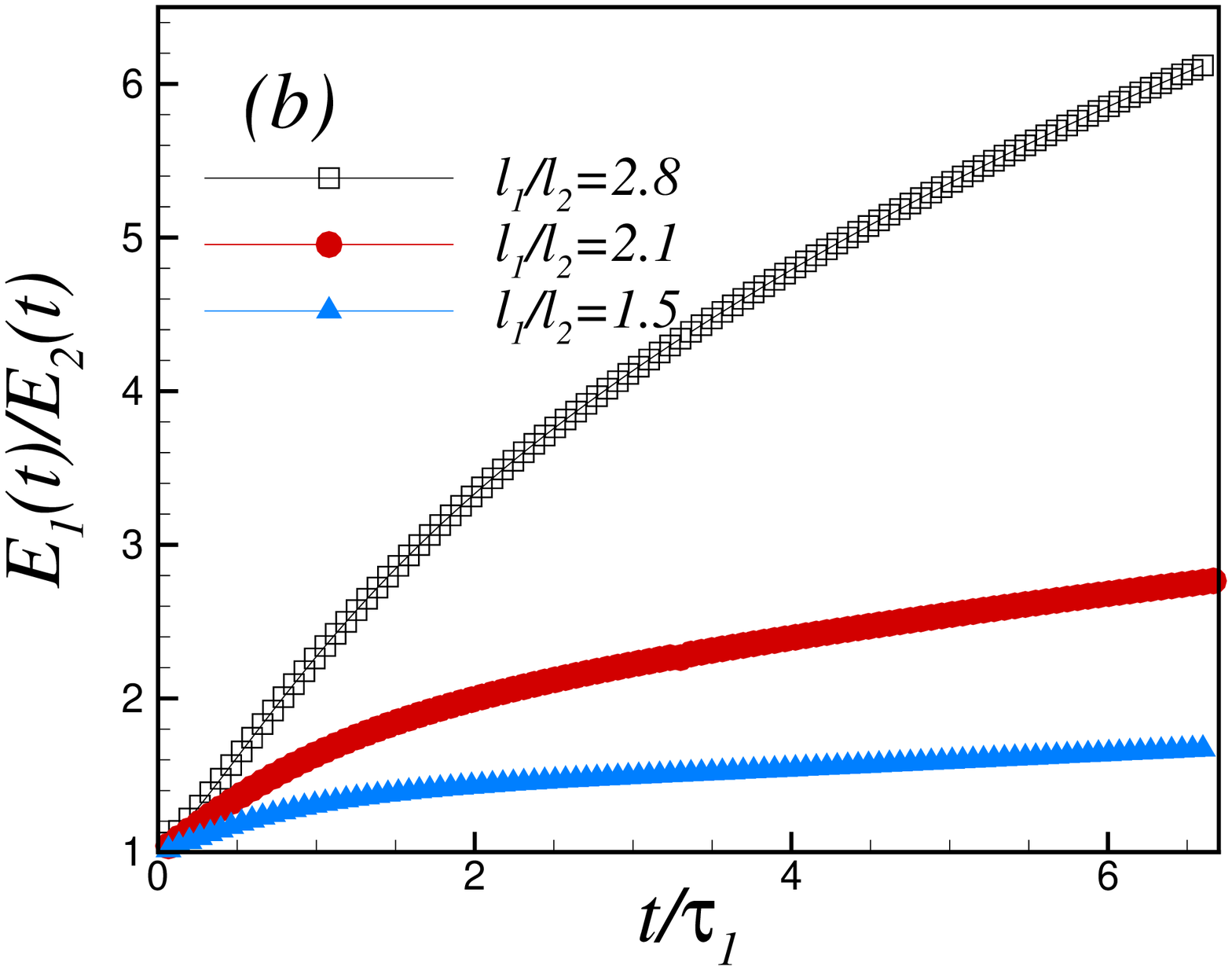}\\
  \vspace*{-7.8cm}\hspace*{0.40\textwidth}\includegraphics[width=0.247\textwidth]{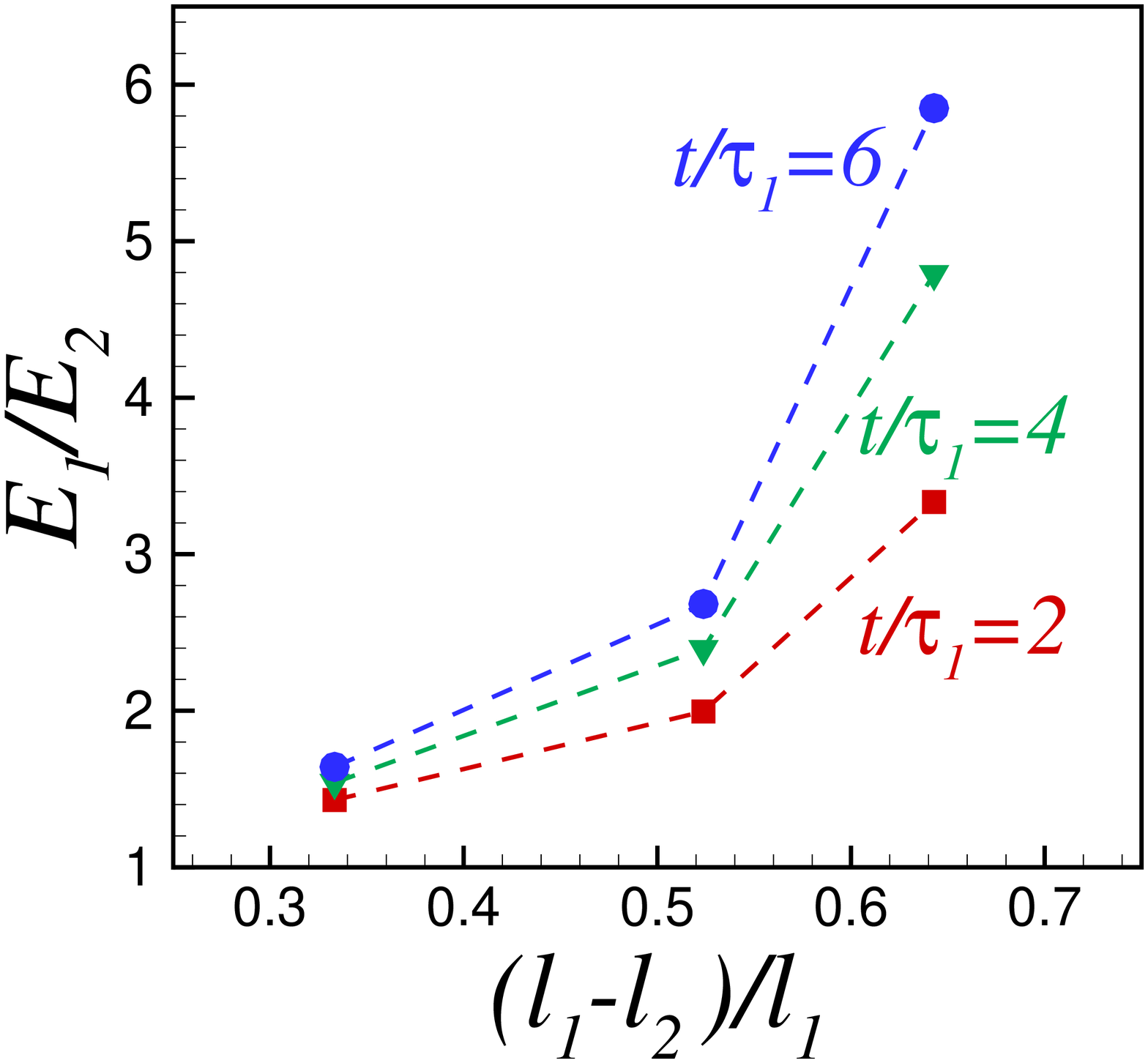}\\[4.5cm]
  \caption{(a) Energy decay of the two initially matched isotropic turbulences: region 1 is the turbulence with the larger correlation length, $Re_\lambda=150$, region 2 is the turbulence with the smaller correlation length. The insert shows the initial Reynolds number (black symbols) and decay exponent (red symbols). (b) Temporal variation of the ratio of the turbulent kinetic energy between the larger and smaller scale flows. The inset shows the energy ratio at three different instants as function of the initial scale gradient imposed across the surface.}
 \label{fig.decay-ratio}
 \end{figure}

\addtolength{\textheight}{1.5cm}
\begin{figure}
\centering
  \vspace*{-15mm}
  \includegraphics[width=0.45\textwidth]{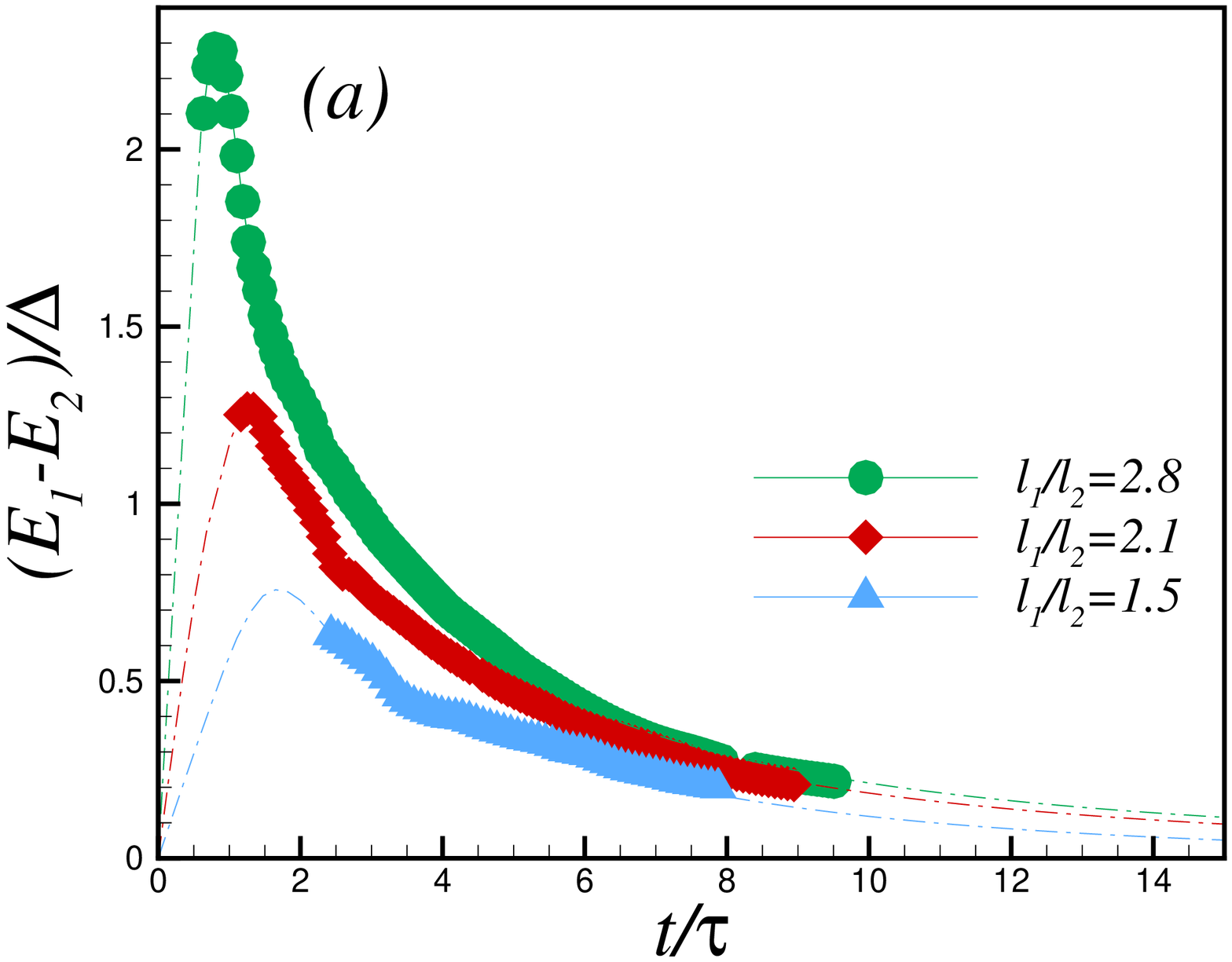}\\
\vspace*{-5.4222cm}\hspace*{0.3022\textwidth}\includegraphics[width=0.2115\textwidth]{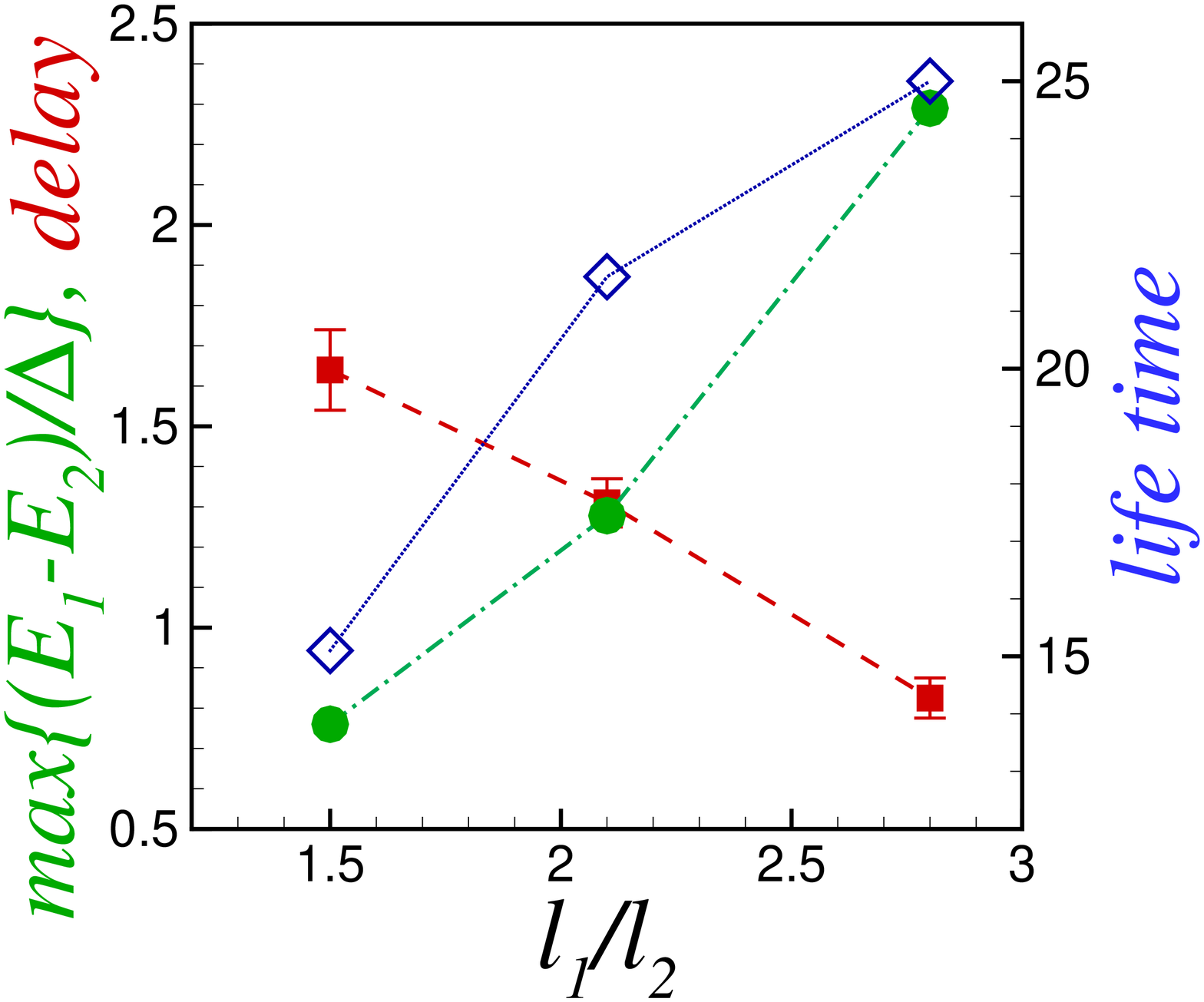}\\[3.4cm]
  \vskip -3mm
  \includegraphics[width=0.45\textwidth]{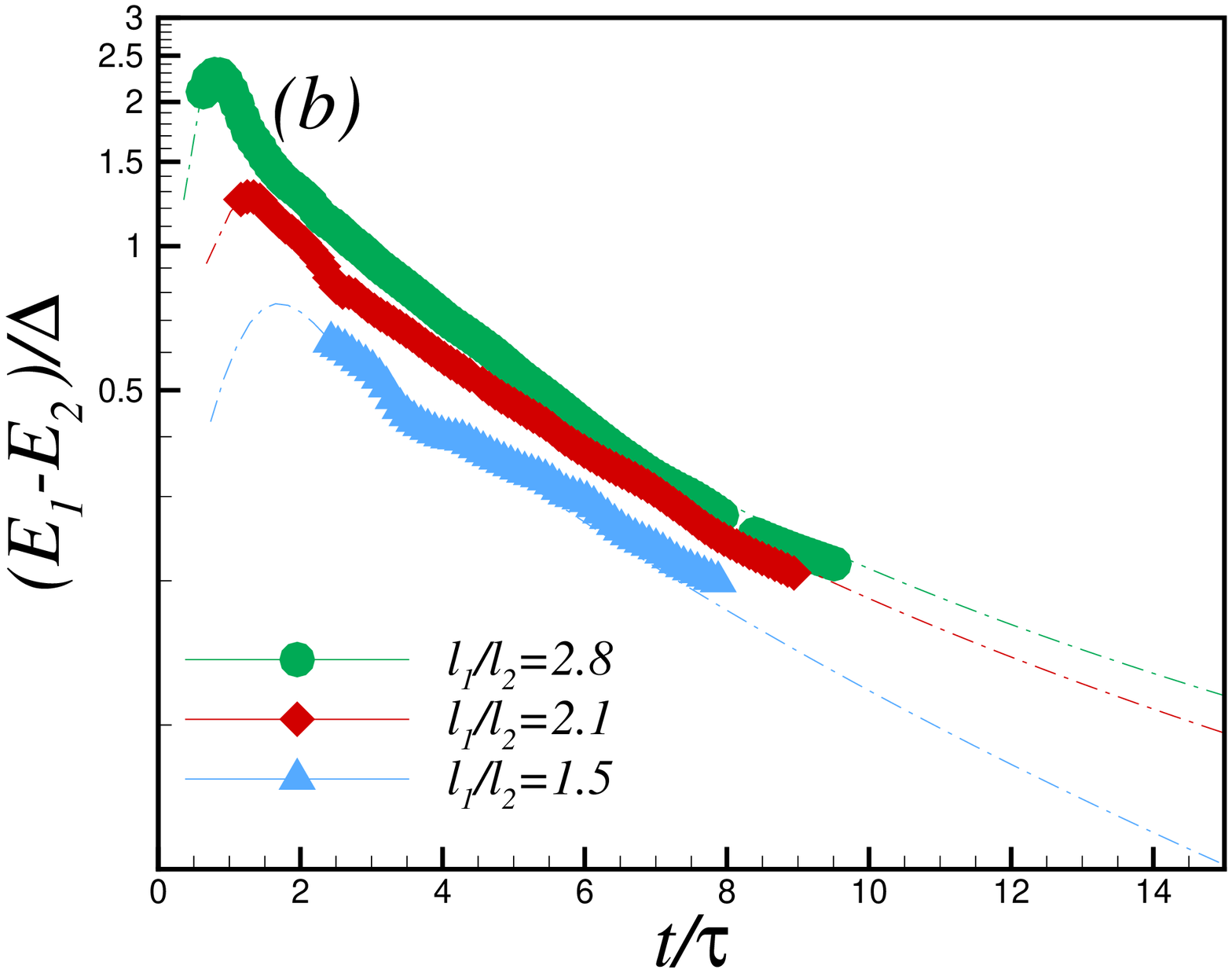}
  \includegraphics[width=0.45\textwidth]{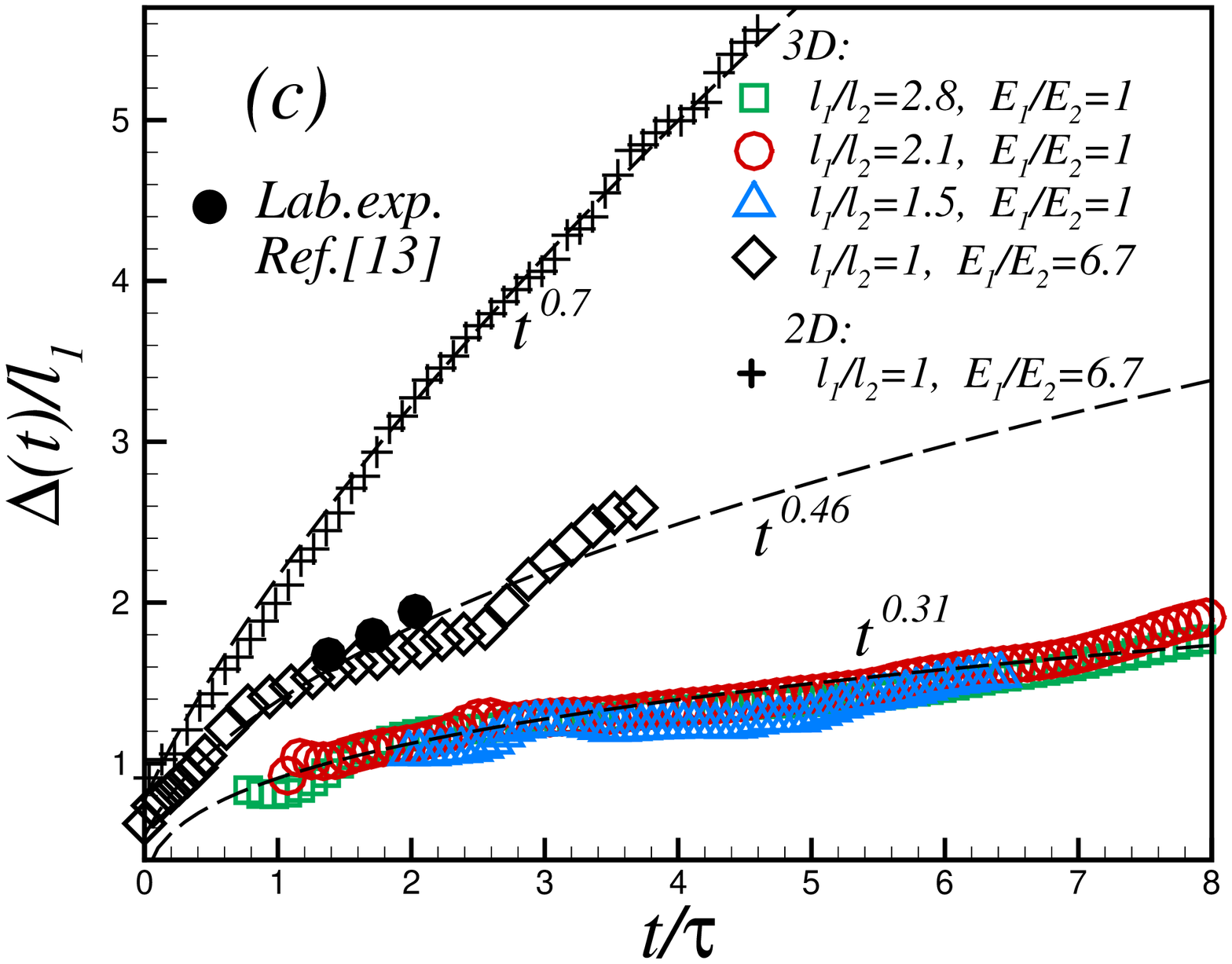}
  \vskip -1mm
  \caption{((a), (b)) Generation of the turbulent kinetic energy gradient. The lines are obtained by the extrapolating the measured data. The inset shows the maximum energy gradient (squares), the delay time at which the maximum of the energy gradient is reached (circles) and the lifetime of the transient (diamonds), defined as the extrapolated time at which the dimensionless energy gradient reduces to 0.05. (c) Time evolution of the mixing length thickness $\Delta$, conventionally defined as the distance between the points where $(E(x,t)-E_1(t))/(E_2(t)-E_1(t))$ has values of 0.75 and 0.25. $E_1$ and $E_2$ are the energies in the two homogeneous regions. Time has been normalized with the average eddy turnover time $\tau=\left(\ell_1+\ell_2\right)/\left(E_1^{1/2} +E_2^{1/2}\right)$. Comparison data from wind tunnel experiments by Veeravalli and Warhaft \cite{vw89} and from two-dimensional simulations \cite{efmc} are also shown.}
 \label{fig.gradient-width}
 \end{figure}


\begin{figure}
 \centering
  \vspace*{-10mm}
  \includegraphics[width=0.48\textwidth]{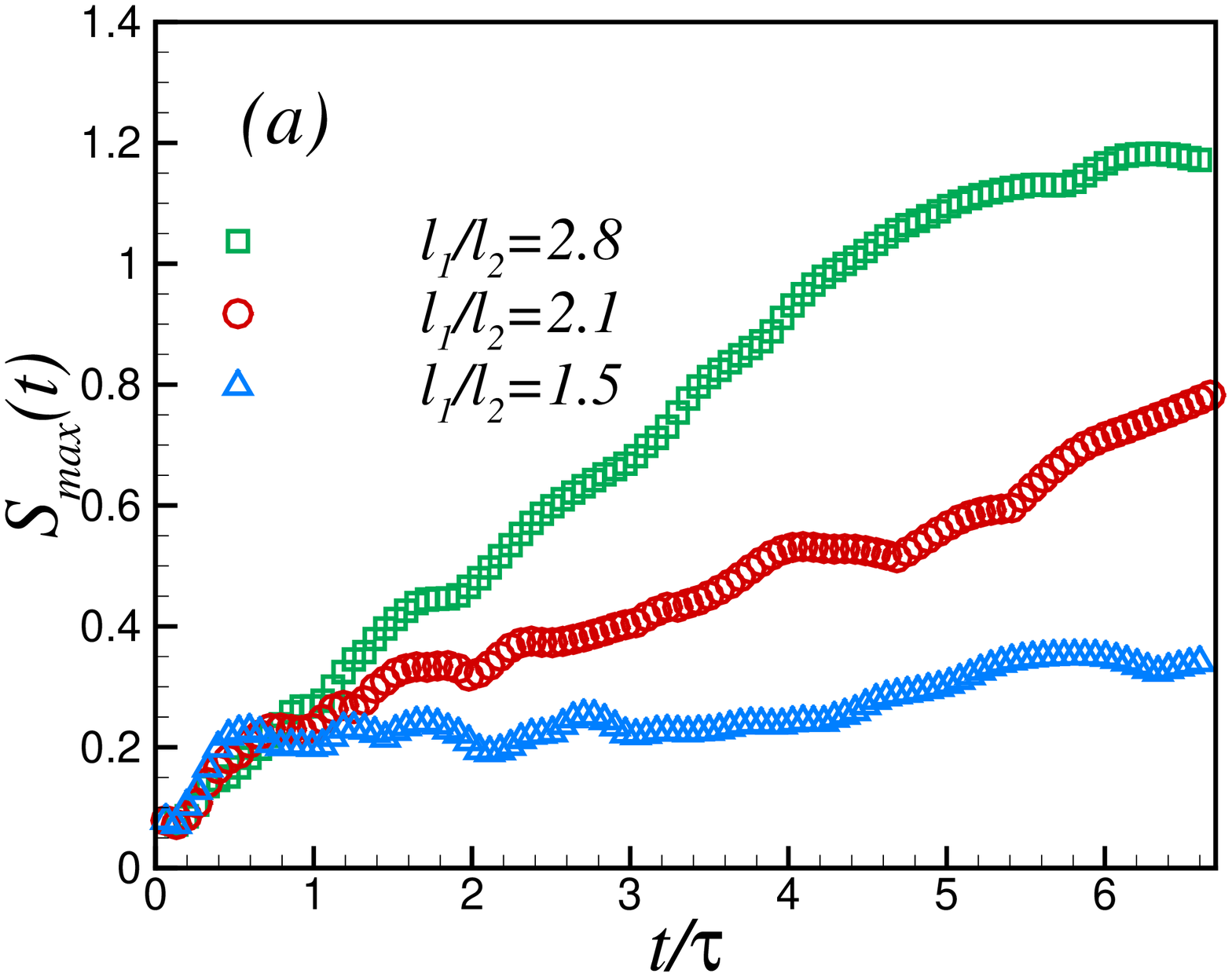}\hfill
  \includegraphics[width=0.48\textwidth]{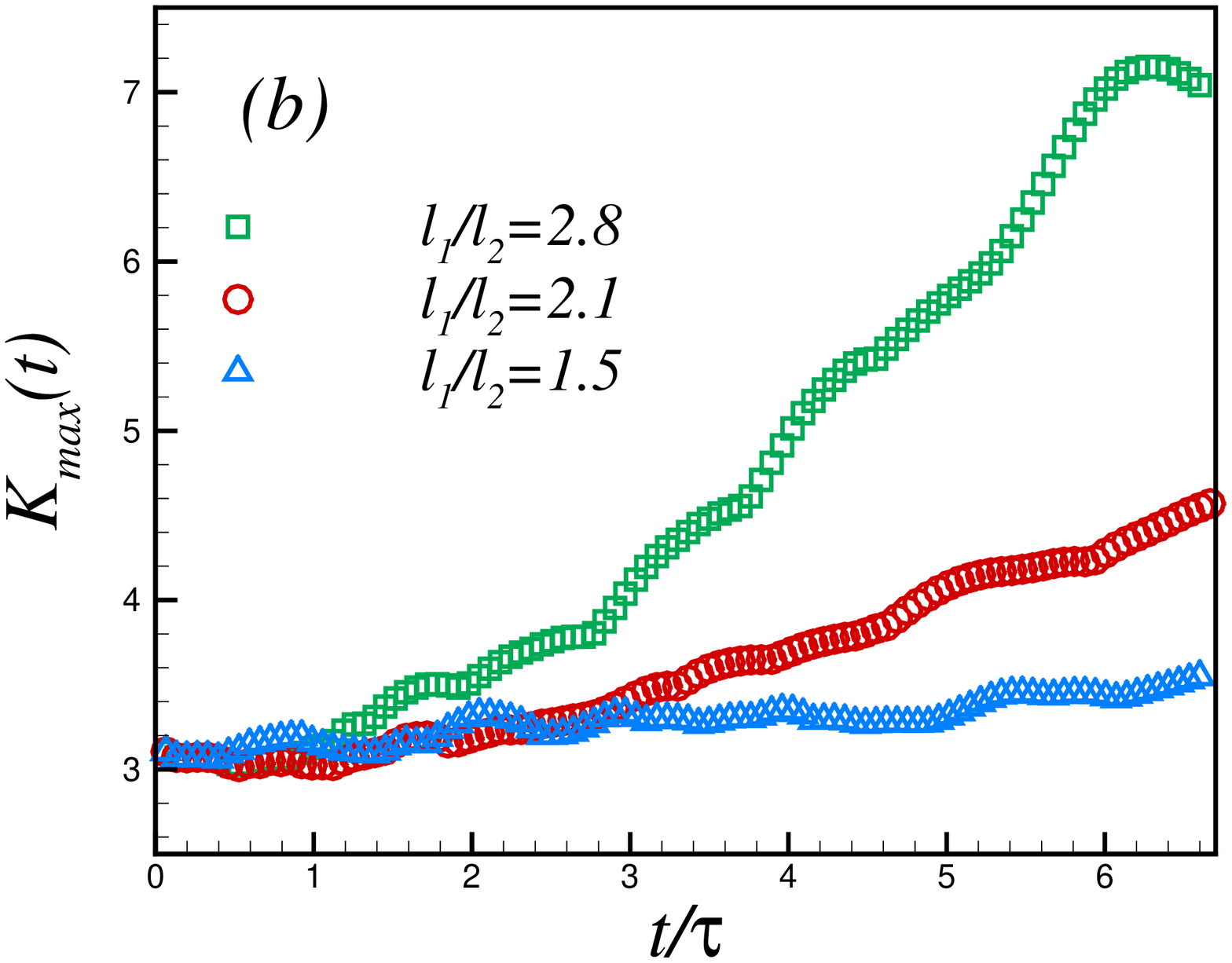}\\[-2.9mm]
  \includegraphics[width=0.48\textwidth]{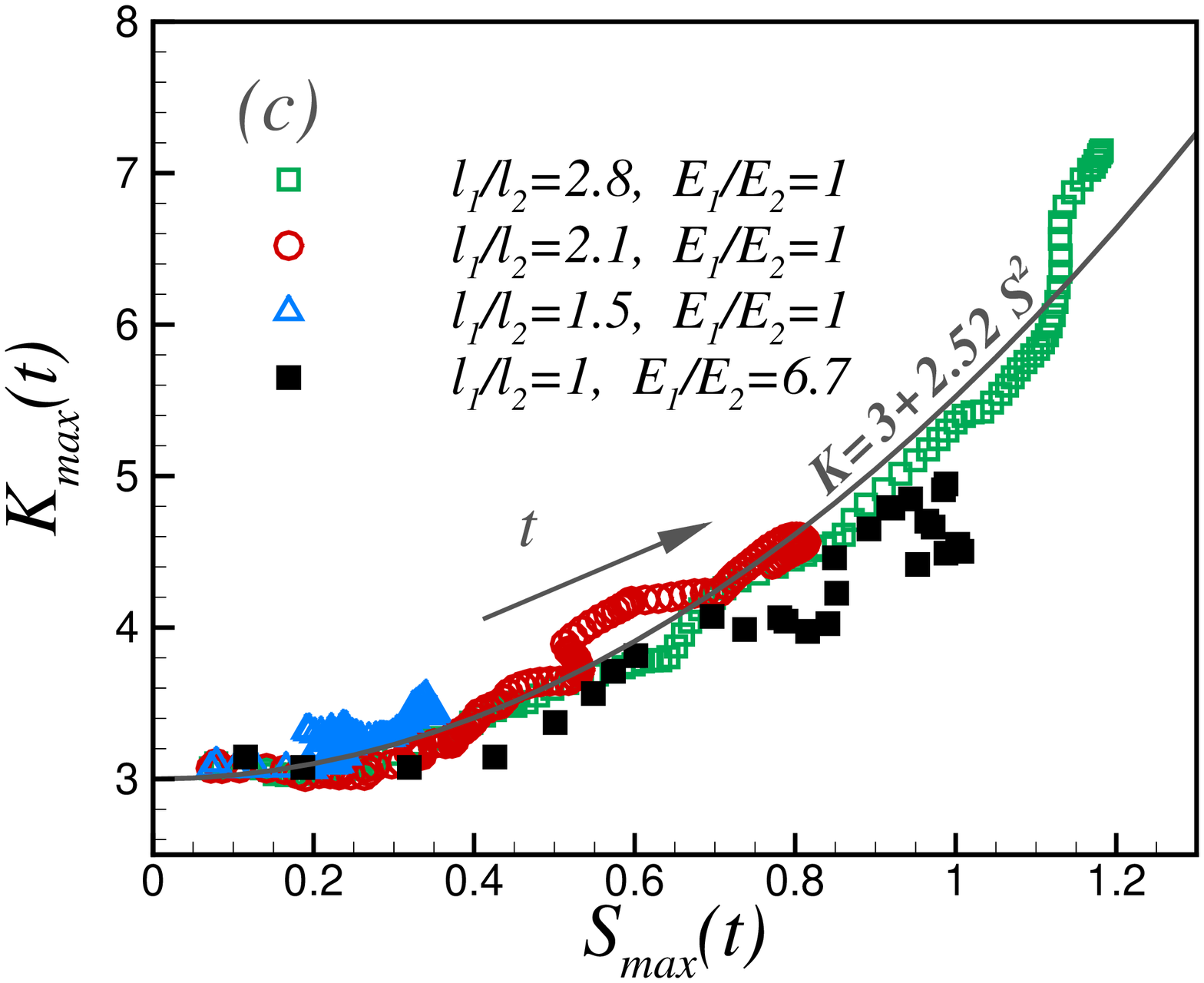}
  \vskip -4mm
  \caption{((a), (b)) \textcolor{black}{Time evolution inside the mixing of the maximum of the skewness $S$ and kurtosis $K$ ($S=\overline{u^3}/(\overline{u^2})^{3/2}$ and $K=\overline{u^4}/(\overline{u^2})^{2}$, where $u$ is the velocity component along the inhomogeneous direction)}.
  (c) Maximum kurtosis as a function of the maximum skewness. The line shows the scaling $K=3+aS^2$ with $a=2.52\pm0.3$ used to fit the whole data set. Data for a mixing with an imposed energy gradient ($E_1/E_2=6.7$) from \cite{ti10} are also shown.}
 \label{fig.sk-vel}
 \end{figure}



\begin{figure}[hbt]
 \hspace*{-2mm}
 \psfrag{S}{\hspace*{-0.4cm}\large$S_{\partial u/\partial x}$}
 \includegraphics[width=0.51\textwidth]{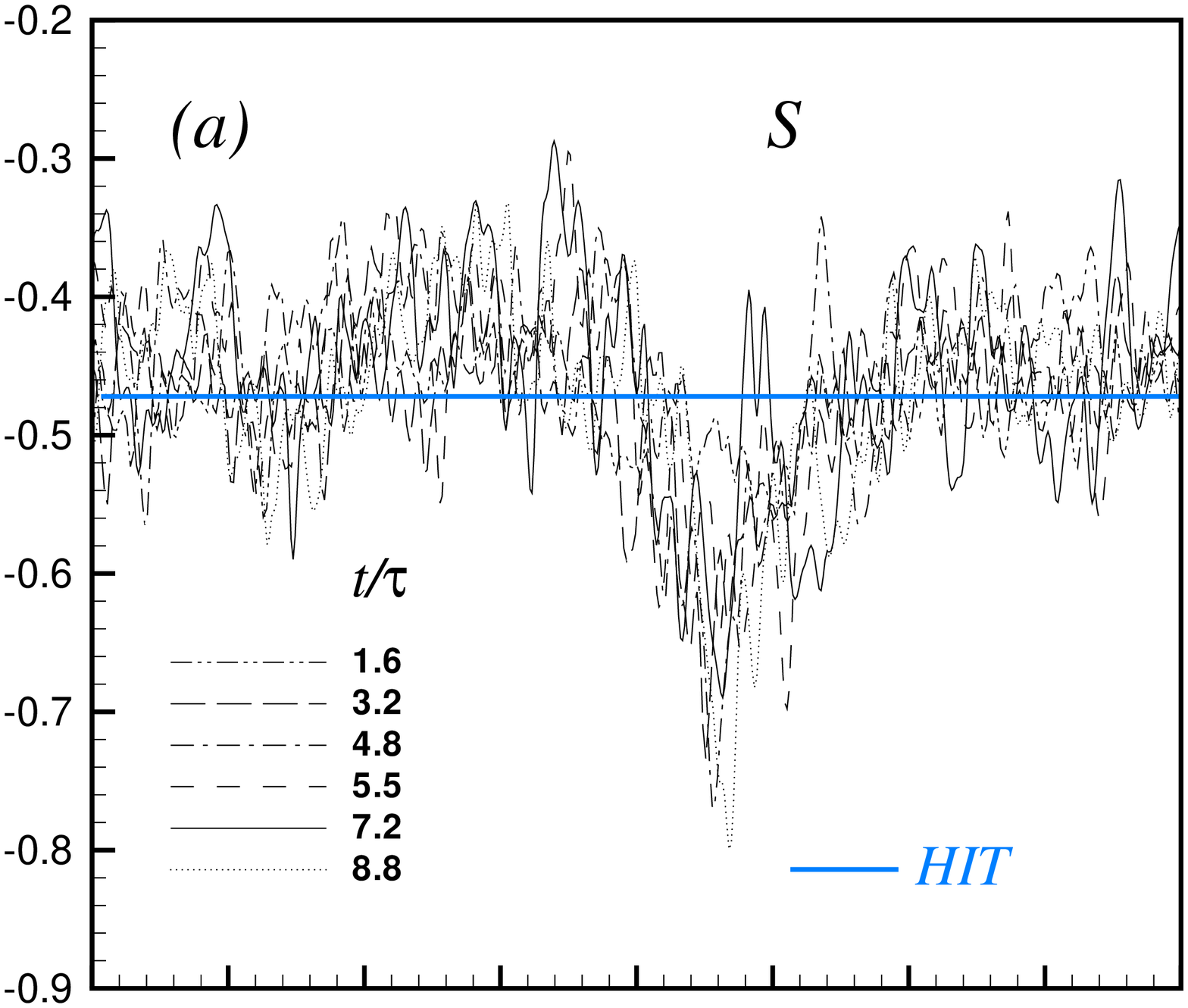}
 \psfrag{S}{\hspace*{-0.4cm}\large$S_{\partial v/\partial y}$}
 \includegraphics[width=0.51\textwidth]{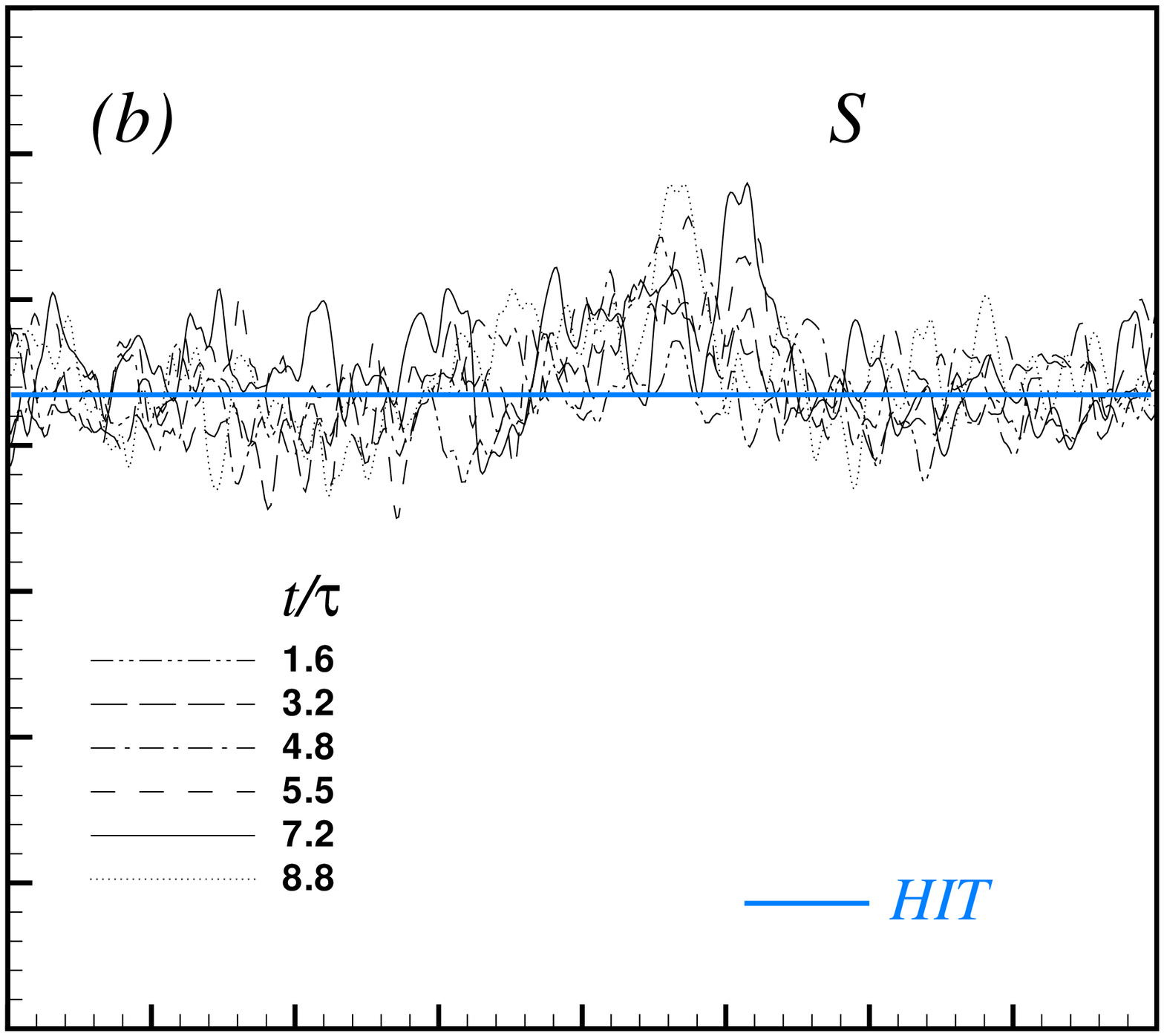}\\
 \hspace*{-2mm}
  \psfrag{K}{\hspace*{-0.4cm}\large$K_{\partial u/\partial x}$}
  \includegraphics[width=0.51\textwidth]{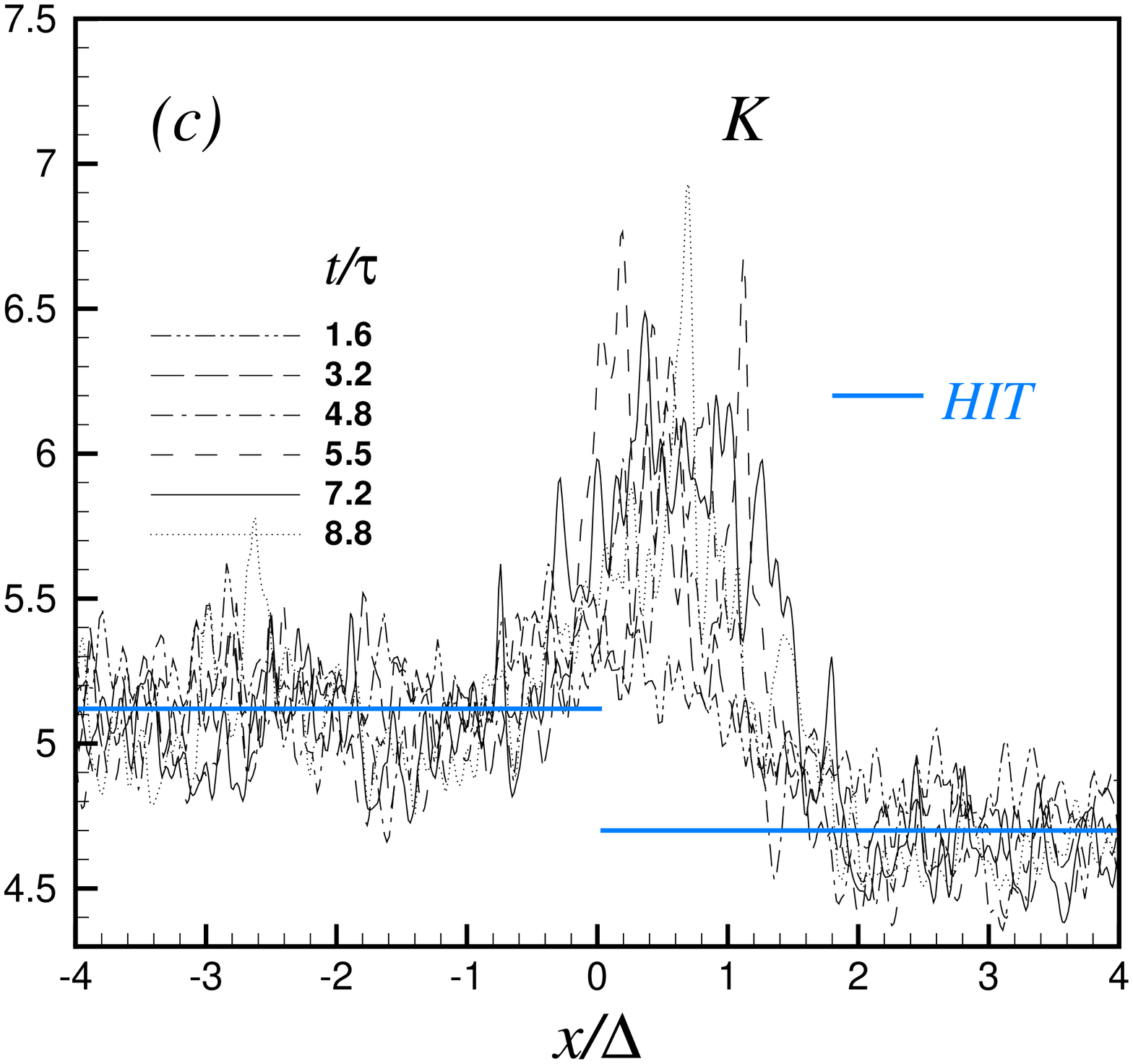} 
  \psfrag{K}{\hspace*{-0.4cm}\large$K_{\partial v/\partial y}$}
  \includegraphics[width=0.51\textwidth]{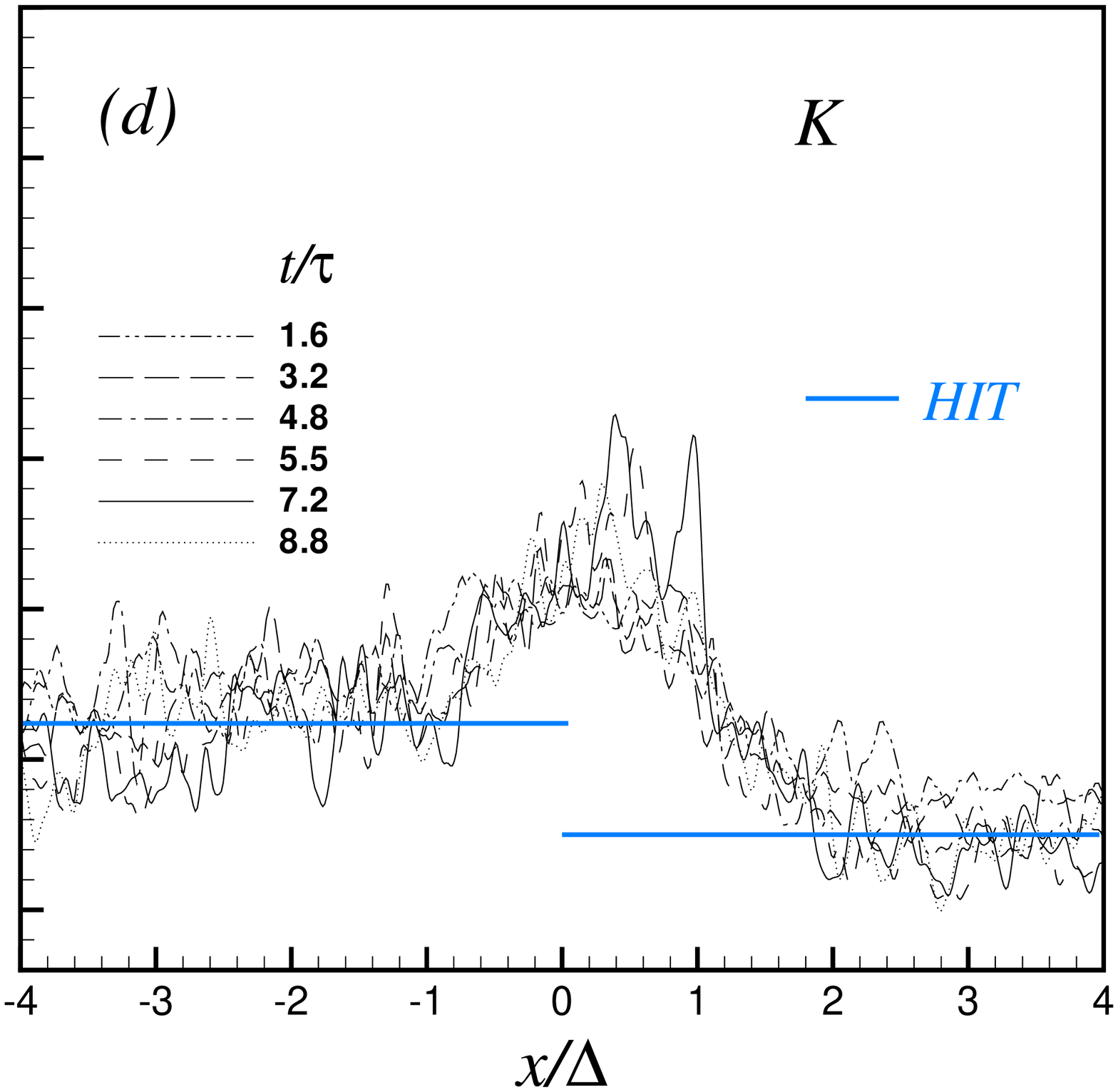}
  \caption{Spatial distribution across the mixing layer of the skewness and kurtosis of the longitudinal derivative $\partial u/\partial x$ in the inhomogeneous direction and $\partial v/\partial y$ in the homogeneous direction from the simulation with an initial scale ratio $\ell_1/\ell_2=2.8$. The horizontal lines indicate the measured values of the derivative moments in the larger-scale homogeneous region ($Re_{\lambda_1}=150$) and in the smaller-scale homogeneous region ($Re_{\lambda_2}=89$).}
  \label{fig.sk-der}
\end{figure}


\begin{figure}[hbt]
\hspace*{-0.05\textwidth}
\centering
\begin{tabular}{l}
 \psfrag{S}{\hspace*{-0.9cm}$S_{\partial u/\partial x}$, $S_{\partial v/\partial y}$}
 \psfrag{R}{\hspace*{-0.9cm}$S_{\partial u/\partial x} / S_{\partial v/\partial y}$}
 \psfrag{L}{\small $\ell_1/\ell_2=1.5$}
 \psfrag{M}{\small $\ell_1/\ell_2=2.1$}
 \psfrag{N}{\small $\ell_1/\ell_2=2.8$}
  \includegraphics[width=0.45\textwidth]{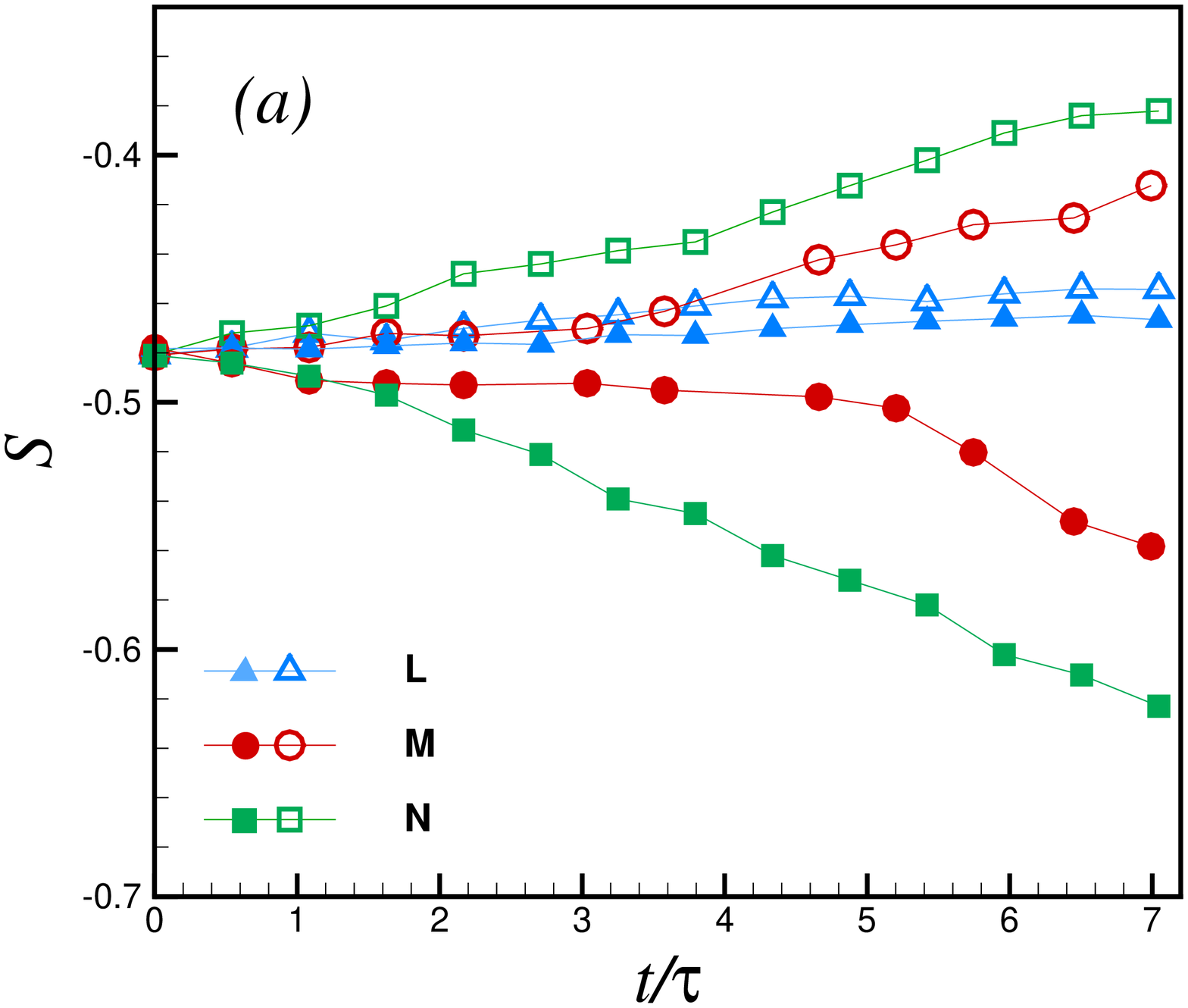}
  \includegraphics[width=0.45\textwidth]{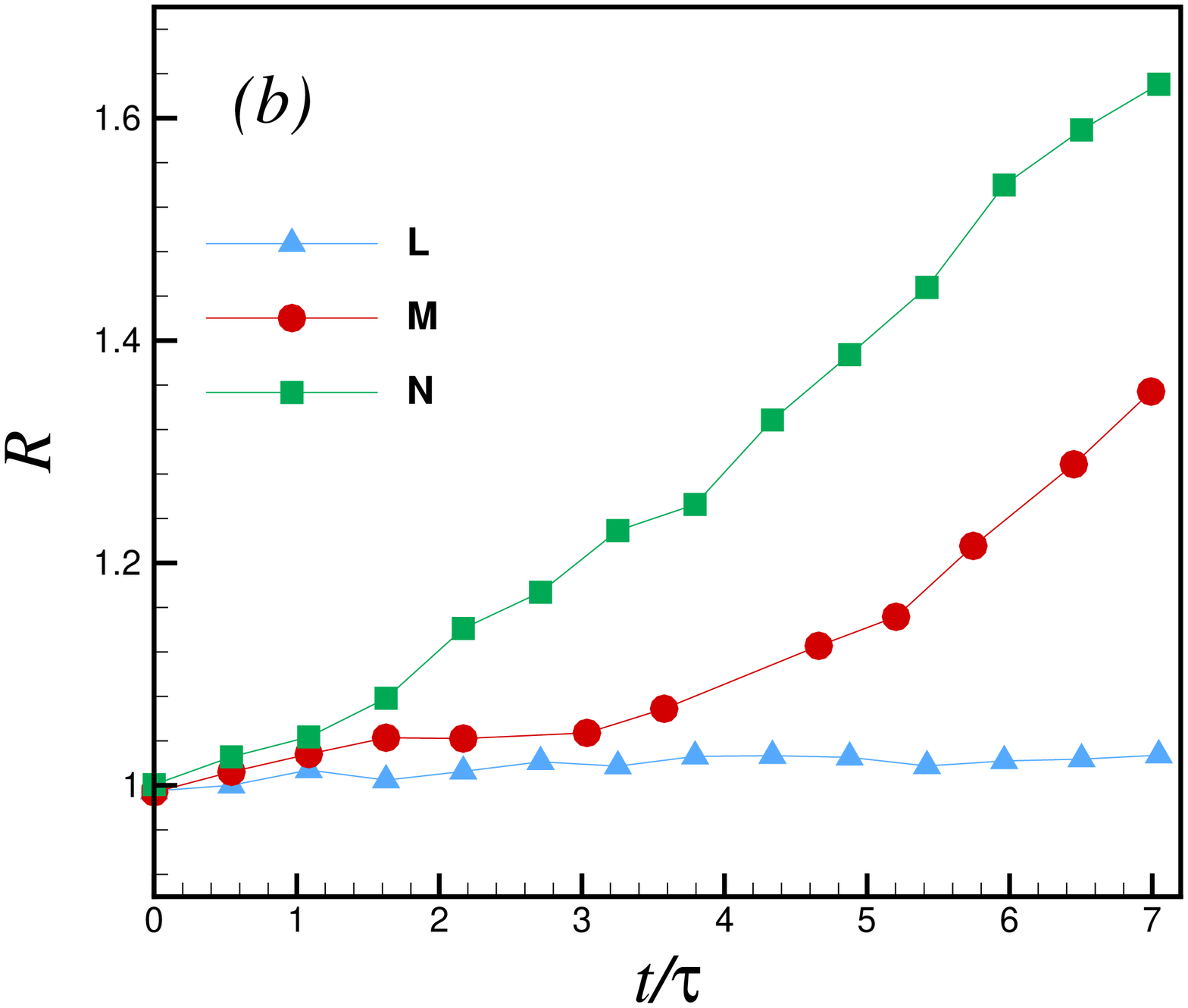}\\ 
 \psfrag{K}{\hspace*{-0.9cm}$K_{\partial u/\partial x}$, $K_{\partial v/\partial y}$}
 \psfrag{R}{\hspace*{-0.9cm}$K_{\partial u/\partial x} / K_{\partial v/\partial y}$}
 \psfrag{L}{\small $\ell_1/\ell_2=1.5$}
 \psfrag{M}{\small $\ell_1/\ell_2=2.1$}
 \psfrag{N}{\small $\ell_1/\ell_2=2.8$}
 \centering
 \psfrag{a}{\small\itshape c}
 \psfrag{b}{\small\itshape d}
  \includegraphics[width=0.45\textwidth]{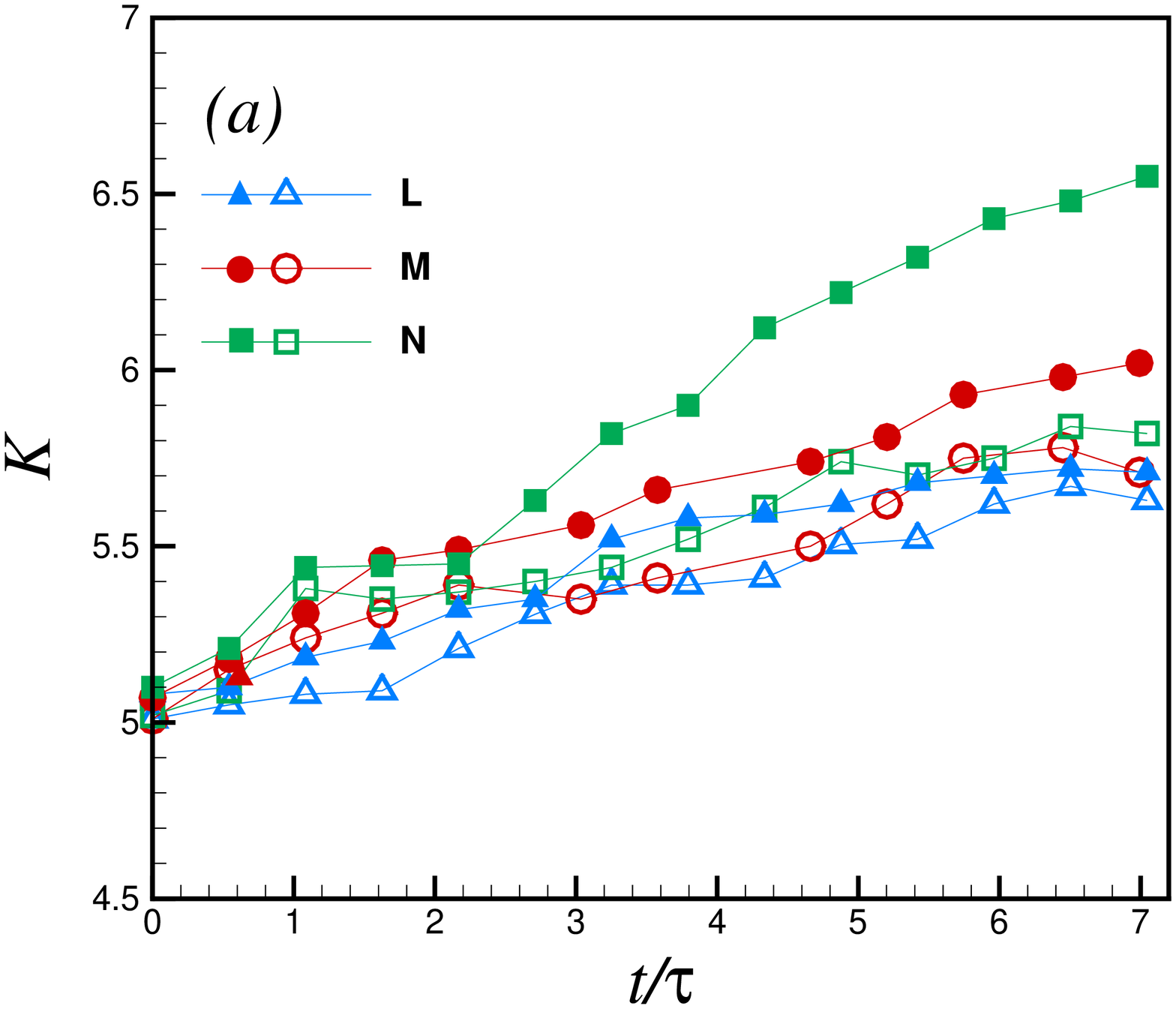}
  \includegraphics[width=0.45\textwidth]{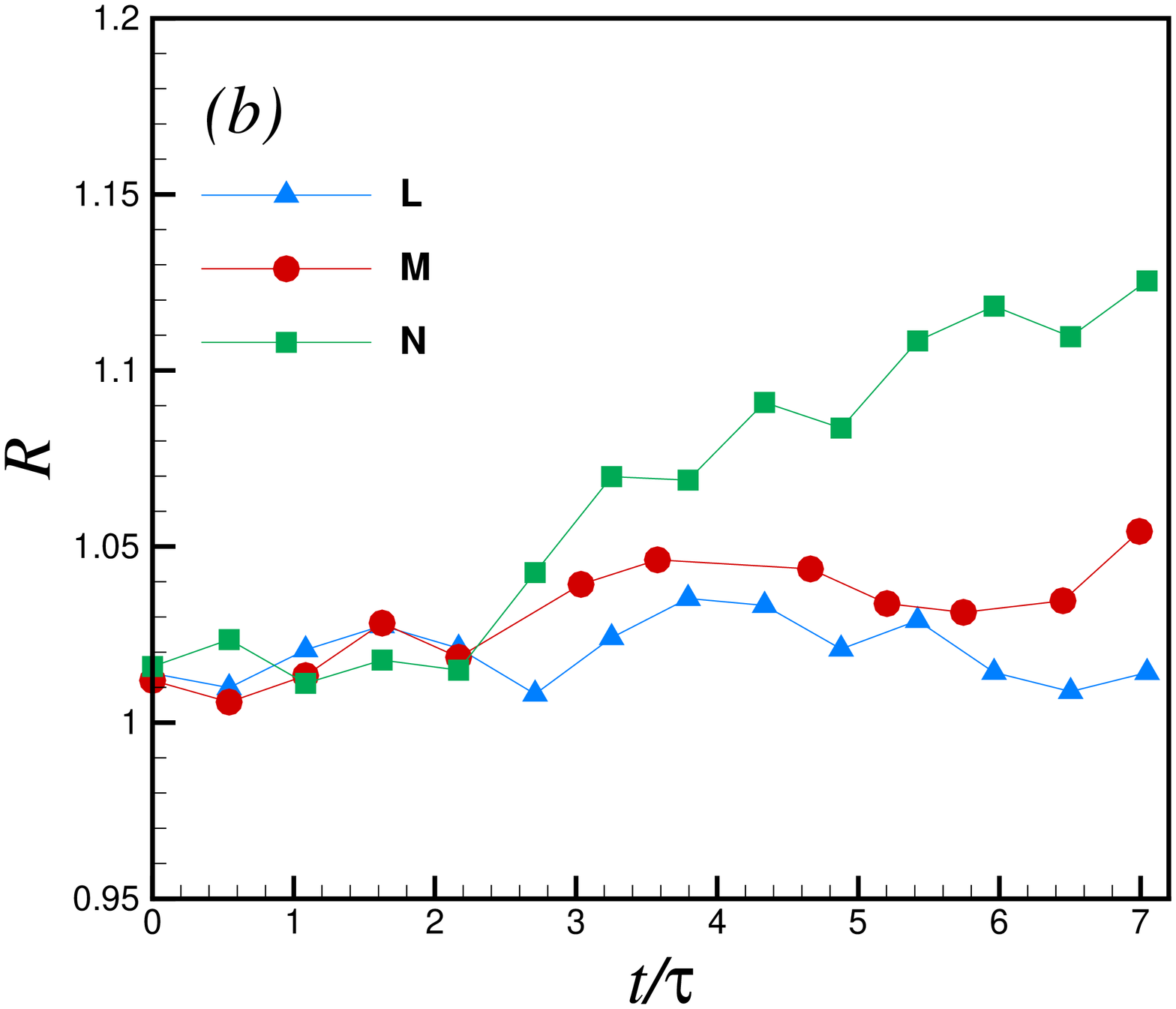}
\end{tabular}\\
\centering
 \psfrag{X}{\hspace*{-1.1cm}$S_{\partial u/\partial x}$, $S_{\partial v/\partial y}$}
 \psfrag{Y}{\hspace*{-1.2cm}$K_{\partial u/\partial x}$, $K_{\partial v/\partial y}$}
 \psfrag{L}{\small $\ell_1/\ell_2=1.5$}
 \psfrag{M}{\small $\ell_1/\ell_2=2.1$}
 \psfrag{N}{\small $\ell_1/\ell_2=2.8$}
\includegraphics[width=0.50\textwidth]{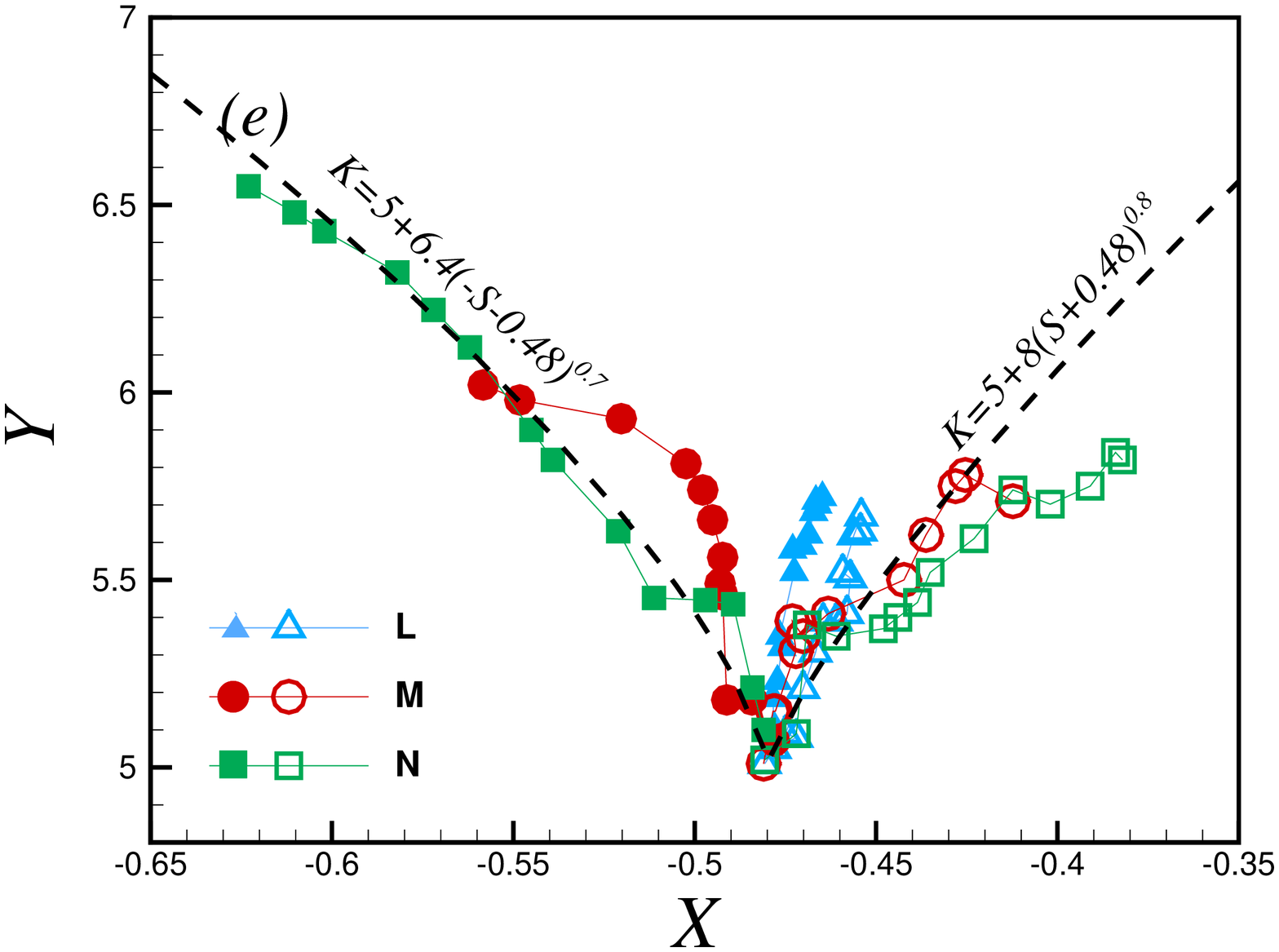}
  \caption{Filled symbols: longitudinal derivative along the inhomogeneous direction $x$, empty symbols: longitudinal derivative along the homogeneous directions $y$. (a-c) Time evolution of the peak of the longitudinal derivative skewness and kurtosis inside the mixing layer. (b-d) Ratio between the longitudinal moments (inhomogeneous over homogeneous directions). (e) Derivative kurtosis as a function of the derivative skewness. The dashed lines show the power law scalings obtained by best fitting. The fitting does not include the data obtained from the simulation with $\ell_1/\ell_2=1.5$, which were not a sufficiently good statistical quality. More runs would be needed to enlarge the statistical base and reduce the noise.}
  \label{fig.peaks-der}
\end{figure}



\section{Method}

The scheme of the flow and of the computational domain can be seen in figure \ref{fig.schema}, part \textit{a}.
The initial conditions are obtained by matching two isotropic fields, obtained from simulations of homogeneous and isotropic turbulence, which have different correlation lengths. These were obtained by using a high pass filter on the velocity field, see \cite{ti06} and the spectra in figure 1{\it c}. Three simulations, with three different ratios of the initial integral scale, 1.5, 2.1 and 2.8, have been carried out. The Reynolds numbers, based on the Taylor microscale of these fields, vary from 70 to 150, with a variation of the correlation length from 0.8 to 2.8 cm \cite{euromech512,ti09}. The matching is made over a layer with an initial width of the same order as the correlation lengths $\ell_1, \ell_2$  \cite{ti06, tib08}. 
The flow is assumed to be contained in a parallelepiped with an aspect ratio of 2 ($4\pi\times(2\pi)^2$) and cyclic boundary conditions are applied to all the spatial directions. The Navier-Stokes equations are solved using a pseudo-spectral Fourier-Galerkin  method with an explicit fourth-order time integration \cite{ict01}. \textcolor{black}{The domain is discretized using 1200 points in the inhomogeneous direction, $x$, and 600 points in the other homogeneous $y_1$ and $y_2$ directions.}

The initial condition is generated as $u_i = u_i^{(1)}p(x)^{\frac{1}{2}} + u_i^{(2)}(1-p(x))^{\frac{1}{2}} +fc(x)$, where $u_i^{(1)}$ and $u_i^{(2)}$ are the two homogeneous and isotropic fields with integral scales $\ell_1$ and $\ell_2$, $p(x)=\left[ 1-\tanh(a\frac{x-L/2}{l_{1}}) \right]/2$ is the matching function and $c(x)f(x)$ is a correction introduced to obtain a uniform energy distribution across the interaction layer, see part \textit{a} of \ref{fig.matching}. This correction is the product of a spatial modulation of the matching region  $c=1-(p^{1/2}+(1-p)^{1/2})$ and a random field $f(x)$ with zero mean and uniform variance $F^2$. The value of $F^2$ is chosen a posteriori in order to generate a uniform energy level $E$ throughout the initial matching layer, see part b) of figure \ref{fig.matching} .
Length  $L$ is the domain size in the $x$ direction and $a$ is a constant, which is chosen in order to have an initial transition layer that is no larger than the integral scale. This kind of correction is not necessary in the experiments where the energy gradient is imposed by the initial conditions \cite{ti06,tib08}.
\textcolor{black}{Field $u^{(1)}$ was obtained from the HIT database by Toschi, see \texttt{http//mp0806.cineca.it/icfd.php}  and also ref.\ \cite{bbcln05}.
Field $u^{(2)}$ has been obtained by filtering out the lowest wavenumbers in $u^{(1)}$ (up to $k=6$); see \cite{ti06}. This field was then been used as the initial condition for a numerical integration of the Navier-Stokes equations which lasts approximately a couple of eddy turnover times and restores the properties relevant to a Navier--Stokes field.} \textcolor{black}{The matched velocity field $u_i$ is then projected onto the solenoidal space in order to enforce incompressibility.}

The  turbulent field $u_i^{(1)}$ has a Taylor microscale Reynolds number equal to 150 and is the same in all the simulations, which differ according to the smaller-scale flow $u_i^{(2)}$. The three-dimensional spectra of the initial conditions are shown in figure \ref{fig.schema}(c). The flows have the same global energy and the different slopes of the large-scale range of the spectra generate different integral scales.
Due to the use of periodic boundary conditions, there are two matching layers in the computational domain, see again  part \textit{a} of figure \ref{fig.schema}.
Directions $y_1$ and $y_2$ in this flow configuration remain statistically homogeneous during the decay, so that
all the statistics can be computed using plane averages in these homogeneous directions. 

\section{Time evolution of the interaction layer}


In this experiment, we have a situation in which it can be observed that the introduction of an inhomogeneity of the correlation length in an otherwise homogeneous turbulence  is sufficient to induce an inhomogeneity in the kinetic energy during the decay. This is a general feature that can be observed in all the present simulations. In figure \ref{fig.schema} (b), it is possible to see the gradual time differentiation of the portions of two flows in the central region of the simulation. In this case, the integral scale ratio is 2.8. The flows with  smaller integral scales decay faster and show higher decay exponents, see figure \ref{fig.decay-ratio}. The decay exponent of the larger-scale flow, whose low wavenumber spectrum has a $k^2$ slope, is -1.15, a result which is close to Saffman's prediction \cite{s67,g92}.  The other decay exponents are -1.3, when the integral scale is reduced by 1.5, -1.46 when it is reduced by 2.1 and -1.65 when is reduced by 2.8, a situation where the low wavenumber spectrum approaches a $k^4$ slope; see \ref{fig.schema}(b) and the inset in figure \ref{fig.decay-ratio} (\textit{a}).
As a consequence, the ratio between the energies in the two regions, initially equal to 1, increases during the whole simulation (figure \ref{fig.decay-ratio} (\textit{b})). During the decay, the large correlation length scale flow also becomes the most energetic one, and an energy gradient, always concurrent with the integral scale gradient, soon emerges, as shown in figure \ref{fig.gradient-width}. The concurrent presence of the gradients, the kinetic energy and the integral scale, in the absence of a mean shear, leads to intermittency and enhances turbulent transport \cite{ti06, tib08}, as discussed in the next section. 

The time evolution of the kinetic energy gradient is plotted in figure \ref{fig.gradient-width} (\textit{a}). This gradient is maximum after about one initial eddy turnover time $\tau=\ell_1/E_1^{1/2}$, which is defined on the initial scales of the field that has the largest correlation length ($E_1$ is the initial turbulent kinetic energy and $\ell_1$ the correlation length). This gradient then gradually decays because the difference between energy levels 1 and 2 also decays in time, see \ref{fig.decay-ratio} (\textit{a}). Furthermore, the interaction length increases, see \ref{fig.gradient-width} (\textit{b}). It can be noticed that the morphology of this transient is very similar to a pattern that is often encountered when observing  three-dimensional  asymptotically stable linear perturbations in sheared  steady flows \cite{c97,stc09,stc10}. Here, however, the lifetime is much shorter than the typical lifetime of hydrodynamic stability, where this kind of transient can last hundreds, even thousands, of base flow timescales. However, at the basic level, the generation-growth-decay-extinction cycle of intermittency can be  considered common in real turbulent flows and representative at a local (zonal) regional level. Local perturbations, associated with more or less intense spatial and temporal variations of the correlation length, are in fact standard features in turbulence. In our opinion, they can be considered analogous to arbitrary perturbations in the hydrodynamic stability theory.


\section{Interaction lifetime, width, intermittency, and anisotropy}

The interaction that we consider can be interpreted as a long-term interaction, since the two isotropic fields are initially separated by a layer which is as thick as the larger correlation length $\ell_1$. The first thing which should be noticed is the  onset of a step variation of the turbulent energy from the initial flat distribution. One is therefore tempted to derive a general notion of  dynamics: a sufficient condition for producing gradients of energy by nonlinear interaction is the presence of variations of the correlation lengths. The mechanism is associated with the different decay exponents that the turbulence shows in the first decade of eddy turnover times on the basis of different correlation lengths. We should recall that the larger the correlation length, the smaller the decay exponent $n$, $E(t) \sim t^n$. A synthesis of the decay properties is presented in figure 3(a) as a function of time and of the correlation scale $\ell$. The inverse relationship between $n$ and the Taylor microscale Reynolds number is shown in the inset. The increase in the energy ratio across the interaction layer  with time and with the gradient of the initial correlation length is presented in figure 3(b). It should be mentioned that, at a constant instant, for instance $t/\tau_1= 6$, a doubling of this scale gradient produces a quadrupling of the energy ratio, see the inset. The effect of this phenomenology is therefore very pronounced, and should be observable and measurable in the laboratory.

The interaction lifetime is finite and can be better described in terms of the energy gradient that starts to be generated at the beginning of the interaction, see figure 4 {\it a} and {\it b}. The gradient intensity is proportional to the ratio (gradient) of the correlation length and the maximum value is reached  in a range of time which is inversely proportional to the gradient intensity. After the peak of the transient evolution is reached, a long decay phase appears, which is due, on one hand, to the growth in the width of the layer, and on the other, to the self-decay of the energy of the two isotropic fields. By extrapolating the instants in which the energy gradient becomes zero, on the basis of the results produced over the first 10 simulation timescales (see in particular the log-linear version of the decay in figure 4(b)), it is possible to see that the lifetime increases almost proportionally to the initial scale ratio; see the inset in figure 4(a). 

This result can be understood and verified by observing the evolution in the diffusion thickness $\Delta$ of the interaction layer. In figure 4 \textit{b}, data from the present simulations are compared with data from a 3D reference simulation over the same computational domain (the energy ratio is here  6.6, with the two isotropic fields sharing the same integral scale \cite{ti10}), with data from  one 2D simulation, with an energy ratio equal to 6.7 and a uniform integral scale \cite{efmc}, and with data from a wind tunnel laboratory experiments \cite{vw89} of shearfree turbulent mixings with a ratio of 6.7. 
The mixing self-diffusion thickness $\Delta$ is conventionally defined as the distance between the points where $(E(x,t)-E_1(t))/(E_2(t)-E_1(t))$ has values of 0.75 and 0.25. $E_1$ and $E_2$ are the energy in the two homogeneous regions. The time has been normalized to the average eddy turnover time $\tau=\left(\ell_1+\ell_2\right)/\left(E_1^{1/2} +E_2^{1/2}\right)$. We can observe that the diffusion lengths of the present simulations collapse very well with this normalization and grow  in a slower manner ($\Delta \sim t ^{0.31}$) than in the case where the initial energy gradient is nonzero. Data related to a case with an energy gradient of about 0.45 J/(kg m), that is, $\ell_1/\ell_2 = 6.7$, are shown in figure 4 \textit{b}.
$\Delta \sim t ^{0.46}$ and  a good agreement between the simulation and laboratory results are observed in three dimensions. There is therefore an evident damping of the turbulent self-diffusion when passing from two-dimensional to three-dimensional dynamics. In the latter case, an efficient driving is the initial presence of the energy gradient; if this is lacking, the mild nonhomogeneity associated with the presence of an integral scale gradient still promotes diffusion, though at a lower rate.
During the transient, the two isotropic flows interact under both the initial macroscale gradient and the newly developed energy gradient, and develop an intermittent sublayer which shows lineaments similar to those  found in shearless mixings, with an initially  imposed energy gradient that has already been discussed in \cite{ti06,tib08,vw89}. 
In two dimensions, where the inverse cascade is active, $\Delta \sim t ^{0.7}$.  It should be noted, that, at present,a comparison with laboratory results is not yet available for either the present or the 2D numerical experiment.

Large-scale intermittency can be observed, looking at the velocity fluctuation $u$ in the $x$ direction, which is the component of the velocity vector that is responsible for the energy transport across the mixing.
\textcolor{black}{By introducing the skewness $S=\overline{u^3}/(\overline{u^2})^{3/2}$ and the kurtosis $K=\overline{u^4}/(\overline{u^2})^{2}$,
we can observe this intermittency through the time evolution of the peak value of $S$ and $K$ inside the mixing layer, see figure \ref{fig.sk-vel}.}
One can observe that the position of this peak
moves, with respect to the centre of $\Delta$, towards the small correlation length side, see \cite{tib08,ti06}. The simulations start from a situation in which the statistics are closer to the isotropic ones, that is, a skewness 0 and a kurtosis close to 3. It is possible to observe the gradual generation of intermittency in fig.\ref{fig.sk-vel}.  It is interesting to note that, when the skewness and kurtosis values are plotted in the $(S,K)$ plane (figure \ref{fig.sk-vel}(c)), all the three-dimensional simulations in part follow the same path, even through they individually present different  growth rates and consequently tend to reach different levels of intermittency. We observe a quadratic relationship $K(S) = 3 + 2.5 S^2$ in this plane. In the case in which the energy gradient drives the interaction from the very  beginning, we can observe an accumulation point ($S\sim7, K\sim4.5$), which can be interpreted as an asymptotic temporal condition.

The intermittent behaviour is not limited to the large scales, i.e. the  velocity moments. In fact, after a few initial eddy turnover times, the longitudinal derivative skewness, which is a normalized dimensionless measure of the average rate of enstrophy generation by vortex stretching, is also increased beyond the typical isotropic turbulence range of values.
Figure \ref{fig.sk-der} shows an example of the spatial distribution of the longitudinal derivative skewness and kurtosis of the velocity component $u$ along the inhomogeneous direction $x$, that is,
$S_{\partial u/\partial x} =
 \overline{\left(\partial u/\partial x\right)^{3}}/
\left(\overline{\left(\partial u/\partial x\right)^2}\right)^{3/2}$
and
$K_{\partial u/\partial x} =
\overline{\left(\partial u/\partial x\right)^{4}}/\left(\overline{\left(\partial u/\partial x\right)^2}\right)^{2}$, in the most intermittent configuration, that is, the one in which the initial integral scale ratio is equal to 2.8.
The most noticeable feature is that the negative value of the longitudinal derivative skewness, whose magnitude is linked to the spectral energy transfer towards small scales, reaches a  magnitude above 0.5, that is, a value which would only be present in isotropic turbulence at much higher Reynolds numbers \cite{sa97,is07,gaw04}. This occurs together with the  increase in intermittency that is visible in the kurtosis distribution.
The longitudinal derivative moments in the directions normal to the interaction layer are also modified. It should be noted that these directions are directions along which the field remains homogeneous. If we look at $S_{\partial v/\partial y} = \overline{\left(\partial v/\partial y\right)^{3}}/\left(\overline{\left(\partial v/\partial y\right)^2}\right)^{3/2}$ and $K_{\partial v/\partial y} = \overline{\left(\partial v/\partial y\right)^{4}}/\left(\overline{\left(\partial v/\partial y\right)^2}\right)^{2}$ (directions $y_1$ and $y_2$ are statistically equivalent), both the skewness and kurtosis show a moderate departure from the values of the neighbouring isotropic regions. We can observe a moderate reduction in the magnitude of the skewness and an increase in the kurtosis (figure \ref{fig.sk-der}).
Such variations of skewness and the derivative are not constant, as could be expected, because the large-scale features that generate the layer are time evolving. Their extreme values are an indication of the departure of the small scale turbulence from the isotropic state, which is observable even in this very mild instance of nonhomogeneity. 

Figures \ref{fig.peaks-der}(a-c) shows the time evolution of these peaks after the fluctuations in the statistics distributions visible in \ref{fig.sk-der} have been filtered out. It is  evident that, in all the configurations, the variation in the derivative skewness in the inhomogeneous ($x$) and homogeneous $y_1, y_2$) directions has an opposite sign: the modulus of $S_{\partial u/\partial x}$ increases, while the modulus of $S_{\partial v/\partial y}$ decreases. The ratio between these two skewness factors increases with time and with the imposed initial scale ratio. For instance, it can be observed that, within about a lapse of about 6-7 timescales, the ratio between the thickness of the longitudinal derivative in the inhomogeneous direction and those in the normal directions (directions along which the field remains homogeneous) increases from 1 to 1.35, when the scale ratio is close to 2, and to 1.64, when the scale ratio is close to 3 (see figure \ref{fig.peaks-der}(b)). This differentiation of the behaviour of the longitudinal derivative, in the energy gradient direction and in the directions normal to it, is qualitatively the same as in the case of the interaction of two isotropic turbulent flows with an imposed energy gradient \cite{ti10}, and seems to be a general feature of an interaction layer in the absence of shear. 
At the same time, the kurtosis levels also differentiate, but even through the inhomogeneous direction shows a slightly higher level of intermittency, the difference is much less marked and never exceeds 30\%. It can be noted that the maximum kurtosis variation is about 25\% of the isotropic value, while the maximum skewness is about 50\% of its isotropic value. The interaction between the two decaying isotropic turbulent flows has a greater impact on the third moment than on the fourth order moment. %
We can conclude that the anisotropy exists at small scales as well as at large scales in the interaction layer between these two flows; the large-scale anisotropy introduced by the initial conditions gradually spreads to the small scales.
However, this kind of anisotropy is different from the kind that can be observed in shear flows, see for example \cite{ws00,ws02}. Here, the main effect is the differentiation of the behaviour of the transverse derivative skewness compared to the longitudinal skewness in the direction of the mean shear (it should be pointed out that those in the normal directions were not been discussed in the previously cited papers). Here, on the contrary, the transport  has a negligible effect on the transverse derivative  moments, which always remain close to their isotropic values.
The conclusion that the postulate of local isotropy is untenable (PLI), both at
dissipation and inertial scales, was also reached in \cite{ws00}. This is because the odd order
moments of the increments of $\delta u(y)$ are nonzero, exhibiting scaling ranges, and the skewness
structure function has a value $\sim 0.5$,  indicating that in the inertial subrange significant anisotropy is
evident even at the third-moment level. Moreover, the fifth-order and seventh-order inertial subrange skewness structure
functions are of the order of 10 and 100, respectively, at least to $Re_{\lambda}$. The approach is usually therefore to check the PLI at various moments with an increase in the control parameter, $ Re_{\lambda}$. In the present work, we approach the PLI
question from a different perspective. We consider the effect of milder and milder nonhomogeneity  on turbulence anisotropy.
For third-order and fourth-order order velocity and velocity derivative moments, we always observe that a substantial degree of anisotropy can be measured.
As a synthesis for the anisotropy structure for the small scales, one can consider the best fitting relations in figure 7{\it e} that link the maxima of the kurtosis and skewness of the velocity derivatives across and along the self-interaction region (note the opposite slopes).

\section{Conclusions}

In conclusion, the numerical experiment that we have performed shows three main results. First, nonlinear interaction is a powerful process that is able to transform a slight inhomogeneity in the correlation length into a kinetic energy step. The interaction can be considered as a long-range one because, initially, the two isotropic fields  are separated by a matching distance which is of the order of the largest integral scale. 

Second,  the interaction builds a sublayer of high intermittency. In fact, the layer which is produced at the interface between the two initial fields  is preceded by this sublayer, which involves not only large scales, but also small scales, and, in particular, their anisotropy. 

Third, the structure of the anisotropy, in this flow configuration, is rather particular. It is easy to observe third-order and fourth-order one-point statistical moments, for either the velocity or velocity derivatives statistics, which follow the structure of the initial condition.
The longitudinal derivative skewness gradually departs from the values that can be expected in an isotropic situation. An appreciable  differentiation can also be observed between the different directions: the skewness in the direction crossing the interaction layer (the direction of inhomogeneity, $x$) increases in magnitude, while it decreases in the direction parallel to the layer. For the kurtosis, the  peak value across the layer is higher for the longitudinal derivative along the inhomogeneous direction. 

It should be noted that the anisotropy described here is different from that of most anisotropic turbulent flows investigated, that is, homogeneous sheared turbulence. In those flows the mean shear creates anisotropy at both large and small scales, which is visible on the generation of large odd transverse moments of the velocity derivative. However, shear has little impact on longitudinal derivatives. In the present situation, the mean velocity is uniform and thus there is no production of turbulent kinetic energy. Departure from local isotropy can be indicated by remarkable difference between the values taken by the longitudinal derivative moments in the directions normal and parallel to the shearfree mixing.

Finally, in this work we have obtained quantitative data on the lifetime and formation timescale of the transients, maximum energy gradient and interaction layer width as a function of the correlation length ratio.

\section*{Acknowledgments}
  The authors would like to thank CINECA and CASPUR computing centres for supporting this work.


\begin{thebibliography}{99}



\bibitem{antonia84}
Anselmet F., Gague Y., Hopfinger E.J.\ and Antonia R.A.
{High-order velocity structure functions in turbulent shear flows},
{\itshape J.\ Fluid Mech.} {\bf 104}, 63--89, (1984).

\bibitem{sa97}
{K.R.\ Sreenivasan , R.A.\ Antonia}
 The phenomenology of small-scale turbulence,
{\it Annual Review of Fluid Mechanics}, {\bf 29}, 435--472 (1997).


\bibitem{aw06}
Ayyalasomayajula S., Warhaft Z.
{Nonlinear interactions in strained axisymmetric high-Reynolds-number
turbulence}, {\it  J.\ Fluid Mech.} {\bf 566}, 273--307, (2006).



\bibitem{bw03}
Bi W.T., Wei Q.D.
{Scaling of longitudinal and transverse structure functions in cylinder wake turbulence},
{\it J.\ Turbul.} {\bf 4}, 028, (2003).

\bibitem{bbcln05}
Biferale L., Boffetta G., Celani A., Lanotte A., Toschi F.
{Particle trapping in three-dimensional fully developed turbulence}
{\it Phys.\ Fluids.} {\bf 17}, 021701/1--4, (2005).

\bibitem{bp05}
Biferale L., Procaccia I.
{Anisotropy in turbulent flows and in turbulent transport},
{\it Phys.\ Rep.} {\bf 414}(2-3), 43--164, (2005).

\bibitem{cps99}
Chertkov M., Pumir A., Shraiman B.I.
{Lagrangian tetrad dynamics and the phenomenology of turbulence},
{\it Phys.\ Fluids} {\bf 11}(8), 2394--2410, (1999).


\bibitem{gfn02}
Gotoh T., Fukayama D., Nakano T.
{Velocity field statistics in homogeneous steady turbulence obtained using a high-resolution direct numerical simulation},
{\it Phys. Fluids} {\bf 14}(3), 1065--1081, (2002).

\bibitem{gaw04}
Gylfason A., Ayyalasomayajula S., Warhaft Z.,
{Intermittency, pressure and acceleration statistics from hot-wire measurements in wind-tunnel turbulence},
{\it J.\ Fluid Mech.} {\bf 501}, 213--229, (2004).



\bibitem{gw98}
Garg S.,Warhaft Z.,
{On the small scale structure of simple shear flow},
{\itshape Phys.\ Fluids} {\bf 10}(3), 662-673, (1998).

\bibitem{ws00}
Warhaft Z., Shen X.,
The anisotropy of the small scale structure in high Reynolds number ($R_\lambda\approx 1000$) turbulent shear flow,
{\it Phys.\ Fluids} {\bf 11}(11), 2976-2989 (2000).

\bibitem{ws02}
Warhaft Z., Shen X.,
On the higher order mixed structure functions in laboratory shear flow,
{\it Phys.\ Fluids} {\bf 14}(7), 2432-2438 (2002).

\bibitem{ssk03}
Schumacher J., Sreenivasan K.R., Yeung P.K.,
Derivative moments in turbulent shear flows,
{\it Phys.\ Fluids} {\bf 15}(1), 84-90 (2003).

\bibitem{w09}
Warhaft Z.,
{Why we need experiments at high Reynolds numbers},
{\it Fluid Dyn.\ Res.} {\bf 41}, 021401/1--10, (2009). 







\bibitem{drazin}
Drazin P.G.,
{\it Introduction to hydrodynamic stability},
Cambridge University press, (2002). 


\bibitem{stc09}
Scarsoglio S., Tordella D., Criminale W.O.,
{An Exploratory Analysis of the Transient and Long-Term Behavior of Small Three-Dimensional Perturbations in the Circular Cylinder Wake},
{\it Studies in Appl.\ Math.} {\bf 123}(2), 153--173, (2009).

\bibitem{stc10}
Scarsoglio S., Tordella D., Criminale W.O.,
{Role of long waves in the stability of the plane wake},
{\it Phys.\ Rev.\ E} {\bf 81}(3), 036326, (2010).


\bibitem{s67}
Saffman P.,
Note on decay of homogenous turbulence.
{\it Phys.\ fluids} {\bf 10}, 1349 (1967).

\bibitem{kd06}
Krogstad P.A., Davidson P.
{Is grid turbulence Saffman turbulence?}
{\it J.\ Fluid Mech.} {\bf 642}, 373­394 (2010).

\bibitem{g92}
{George W.K.},
 {The decay of homogeneous isotropic turbulence.}
{\it Phys.\ Fluids} {\bf 4}, 1492-1509 (1992).


\bibitem{sb92}
{Speziale C.G., Bernard P.},
{The energy decay in self-preserving isotropic turbulence revisited.}
{\it J. Fluid Mech.} {\bf 241}, 645--667 (1992).


\bibitem{gd04}
{George W.K., Davidson L.},
{Role of initial conditions in establishing asymptotic flow behavior.}
{\it AIAA J.}\ {\bf 42}, 438-446 (2004).

\bibitem{ss00}
Skrbeka L., Stalp R.,
On the decay of homogeneous isotropic turbulence
{\it Phys.\ Fluids} {\bf 12}(8), 1997-2019 (2000).

\bibitem{a07}
{ Lavoie P., Djenidi L., Antonia R.A.},
 {Effects of initial conditions in decaying turbulence generated by passive grids.}
{\i J.Fluid Mech.}\ {\bf 585}, 395-420 (2007).



\bibitem{euromech512}
Tordella D., Iovieno M.,
{Small scale anisotropy induced by a spatial variation of the integral scale},
{\itshape Atti dell'Accademia delle Scienze di Torino. Classe di Scienze Fisiche, Matematiche e Naturali}, {\bf 142}, suppl.2009, 108-111, (2008), proceedings of the Euromech Colloquium 512 ``Small Scale Turbulence and Related Gradient Statistics''.


\bibitem{ti09}
{Tordella D., Iovieno M.},
 {Step onset from an initial uniform distribution of turbulent kinetic energy}, {\it Advances in Turbulence XII}, 677-680, Springer (2009). Proceedings of the {\it 12th European Turbulence Conference}, Marburg, September 7-10, 2009.

\bibitem{ti06}
{ Tordella D., Iovieno M.},
 {Numerical experiments on the intermediate asymptotics of the free shear turbulence mixing.}
{\it J.\ Fluid Mech.}\ {\bf 549}, 441-454  (2006).


\bibitem{tib08}
{ Tordella D., Iovieno M., Bailey P.R.}
 {Sufficient condition for Gaussian departure in turbulence},
{\it Phys.\ Rev.\ E}, \textbf{77}, 016309 (2008).


\bibitem{ict01}
{Iovieno M., Cavazzoni C., Tordella D.}
{A new technique for a parallel dealiased pseudospectral Navier-Stokes code.}
{\it Comp.\ Phys.\ Comm.}, {\bf 141}, 365--374, (2001).


\bibitem{c97}
Criminale W.O., Jackson T.L., Lasseigne D.G., Joslin R.D.,
{Perturbation dynamics in viscous channel flows}
{\it J.\ Fluid Mech.} {\bf 339}, 55--75, (1997).


\bibitem{is07}
Ishihara T., Kaneda Y., Yokokawa M., Itakura K., Uno A. 
Small-scale statistics in high-resolution direct numerical simulation of turbulence: Reynolds number dependence of one-point velocity gradient statistics,
{\it J.Fluid Mech.} {\bf 595}, 335-366 (2007).













\bibitem{vw89}
{ Veeravalli S., Warhaft Z.},
 {The shearless turbulence mixing layer.}
{\it J.\ Fluid Mech.}\ {\bf 207}, 191-229, (1989).





%




\bibitem{ti10}
Tordella D., M.Iovieno M.,
Small scale anisotropy in turbulent shearless mixing,
under review for {\itshape Phys.\ Rev.\ Lett.}, (2010).


\bibitem{efmc}
Iovieno M., Ducasse L., Tordella D.,
Passive scalar diffusion through an energy step,
{\itshape 8th European Fluid Dynamics Conference}, Bad Reichenhall, 13-16 September (2010).



\end{thebibliography}
\end{document}